\documentclass[11pt]{article}

\usepackage[T1]{fontenc}
\usepackage[utf8]{inputenc}
\usepackage{lmodern}
\usepackage{textcomp}
\usepackage{amsmath,amssymb,amsfonts}
\usepackage{bm}
\usepackage[a4paper,margin=1in]{geometry}
\usepackage{xcolor}
\usepackage{graphicx}
\usepackage{longtable,booktabs,array}
\usepackage{calc}
\usepackage{etoolbox}
\usepackage{microtype}
\usepackage[normalem]{ulem}
\usepackage{parskip}
\usepackage{newunicodechar}
\usepackage[colorlinks=true,allcolors=blue!55!black,breaklinks=true]{hyperref}
\usepackage{xurl}

\makeatletter
\patchcmd\longtable{\par}{\if@noskipsec\mbox{}\fi\par}{}{}
\makeatother

\makeatletter
\def\maxwidth{\ifdim\Gin@nat@width>\linewidth\linewidth\else\Gin@nat@width\fi}
\def\maxheight{\ifdim\Gin@nat@height>\textheight\textheight\else\Gin@nat@height\fi}
\makeatother
\setkeys{Gin}{width=\maxwidth,height=\maxheight,keepaspectratio}

\newenvironment{refslist}%
  {\par\small\sloppy\setlength{\parindent}{0pt}\setlength{\parskip}{4pt plus 1pt}%
   \everypar{\hangindent=1.5em\hangafter=1}}%
  {\par}

\newunicodechar{ }{\,}
\newunicodechar{‐}{-}
\newunicodechar{–}{--}
\newunicodechar{“}{``}
\newunicodechar{”}{''}
\newunicodechar{•}{\textbullet{}}
\newunicodechar{ˉ}{\ifmmode\bar{}\else\textasciimacron\fi}
\newunicodechar{°}{\textdegree{}}
\newunicodechar{±}{\ensuremath{\pm}}
\newunicodechar{×}{\ensuremath{\times}}
\newunicodechar{·}{\ensuremath{\cdot}}
\newunicodechar{⋅}{\ensuremath{\cdot}}
\newunicodechar{½}{\textonehalf{}}
\newunicodechar{√}{\ensuremath{\surd}}
\newunicodechar{∝}{\ensuremath{\propto}}
\newunicodechar{∼}{\ensuremath{\sim}}
\newunicodechar{≈}{\ensuremath{\approx}}
\newunicodechar{≠}{\ensuremath{\neq}}
\newunicodechar{≥}{\ensuremath{\geq}}
\newunicodechar{−}{\ensuremath{-}}
\newunicodechar{∂}{\ensuremath{\partial}}
\newunicodechar{∇}{\ensuremath{\nabla}}
\newunicodechar{′}{\ensuremath{{}^{\prime}}}
\newunicodechar{↑}{\ensuremath{\uparrow}}
\newunicodechar{↓}{\ensuremath{\downarrow}}
\newunicodechar{→}{\ensuremath{\rightarrow}}
\newunicodechar{↔}{\ensuremath{\leftrightarrow}}
\newunicodechar{¹}{\ensuremath{^{1}}}
\newunicodechar{²}{\ensuremath{^{2}}}
\newunicodechar{³}{\ensuremath{^{3}}}
\newunicodechar{⁴}{\ensuremath{^{4}}}
\newunicodechar{⁵}{\ensuremath{^{5}}}
\newunicodechar{⁶}{\ensuremath{^{6}}}
\newunicodechar{⁷}{\ensuremath{^{7}}}
\newunicodechar{⁹}{\ensuremath{^{9}}}
\newunicodechar{⁺}{\ensuremath{^{+}}}
\newunicodechar{⁻}{\ensuremath{^{-}}}
\newunicodechar{₀}{\ensuremath{_{0}}}
\newunicodechar{₁}{\ensuremath{_{1}}}
\newunicodechar{₂}{\ensuremath{_{2}}}
\newunicodechar{Γ}{\ensuremath{\Gamma}}
\newunicodechar{Δ}{\ensuremath{\Delta}}
\newunicodechar{Θ}{\ensuremath{\Theta}}
\newunicodechar{α}{\ensuremath{\alpha}}
\newunicodechar{β}{\ensuremath{\beta}}
\newunicodechar{γ}{\ensuremath{\gamma}}
\newunicodechar{δ}{\ensuremath{\delta}}
\newunicodechar{η}{\ensuremath{\eta}}
\newunicodechar{κ}{\ensuremath{\kappa}}
\newunicodechar{λ}{\ensuremath{\lambda}}
\newunicodechar{μ}{\ensuremath{\mu}}
\newunicodechar{µ}{\ensuremath{\mu}}
\newunicodechar{σ}{\ensuremath{\sigma}}
\newunicodechar{τ}{\ensuremath{\tau}}
\newunicodechar{φ}{\ensuremath{\phi}}
\newunicodechar{ϕ}{\ensuremath{\phi}}
\newunicodechar{ψ}{\ensuremath{\psi}}
\newunicodechar{á}{\'a}
\newunicodechar{ä}{\"a}
\newunicodechar{è}{\`e}
\newunicodechar{é}{\'e}
\newunicodechar{ö}{\"o}
\newunicodechar{ü}{\"u}
\newunicodechar{ß}{\ss{}}
\newunicodechar{Ă}{\u{A}}
\newunicodechar{ă}{\u{a}}

\begin{document}

\begin{center}
{\LARGE\bfseries Neuronal electricality founded in murburn-thermodynamic principles}\\[5pt]
{\large\bfseries Comparisons, evidenced explanations, and predictions}\\[6pt]
{\large \emph{Kelath Murali Manoj\textsuperscript{1,2}, N Sukumar\textsuperscript{1}, Taufia Hussain}\textsuperscript{3}\emph{, Mahendra Kavdia}\textsuperscript{4}\emph{, Abhijith Anandakrishnan}\textsuperscript{1}}\\[8pt]
{\footnotesize \textsuperscript{1} Amrita School of Artificial Intelligence, Coimbatore, Amrita Vishwa Vidyapeetham, Amritanagar, Ettimadai 641112, Tamil Nadu, India. \\[2pt]
\textsuperscript{2} Satyamjayatu: The Science \& Ethics Foundation, Shoranur-2, Palakkad Dist. 679122, Kerala, India. \\[2pt]
\textsuperscript{3}Department of Animal Physiology, Institute of Biosciences, Albert-Einstein-Straße 3, University of Rostock, 18059 Rostock, Germany \\[2pt]
\textsuperscript{4}Department of Biomedical Engineering, Wayne State University, \\[2pt]
Detroit, Michigan, USA \\[2pt]
*KMM: km\_manoj@cb.amrita.edu; murman@satyamjayatu.com; ORCID: 0000-0003-4515-994X \\[2pt]
NS: n\_sukumar@cb.amrita.edu; ORCID: 0000-0002-2724-9944 \\[2pt]
TH: taufia.hussain@uni-rostock.de; ORCID: 0009-0000-8334-4762 \\[2pt]
MK: kavdia@wayne.edu \\[2pt]
AA: a\_abhijith@cb.amrita.edu; ORCID: 0000-0002-4650-8438}
\end{center}

\begin{abstract}
\noindent The analyses presented herein demonstrate that neuronal electrical activity can be consistently interpreted as a manifestation of murburn redox-mediated electronic dynamics rather than as a process fundamentally driven by transmembrane ionic flux. By integrating comparison with established models, quantitative predictions, and diverse experimental observations, the murburn framework emerges as a unified and chemically grounded description of excitability. A key strength of the model lies in its predictive structure. Its appeal is mechanistic parsimony rather than freedom from fitting (the model still uses free or imposed parameters); the murburn formulation links measurable electrophysiological outputs: such as conduction velocity, waveform morphology, and threshold behavior; to physically interpretable variables including redox kinetics, transport efficiency, and environmental conditions. This enables direct experimental validation through perturbations in oxygen availability, redox balance, solvent properties, ionic strength, and external fields. Importantly, the framework extends beyond neurons to a broader class of excitable systems, including cardiac tissue, photoreceptors, and artificial redox-active materials, suggesting that excitability is a general physicochemical phenomenon rooted in reaction-transport dynamics. While the present work establishes the mid-scale dynamics of neuronal electricality, further developments are required to connect quantum-level electron transfer processes with macroscopic electrophysiological signals such as EEG and EMG. These extensions, along with targeted experimental tests, will determine the ultimate scope and applicability of the murburn paradigm.

\end{abstract}

\section*{1. Introduction}

The first part of this study introduced the background of electrophysiology, presented a murburn-based physicochemical framework for neuronal electrical activity and derived the governing master equation for electron-holding potential dynamics. The present work extends that theoretical foundation by examining the explanatory scope, comparison with other systems/models, lays out experimental predictions, and details the physiological implications of the murburn redox model. Specifically, we compare the predictions of the murburn framework with those of classical channel--pump electrophysiology across a wide range of perturbations including temperature, solvent properties, ionic strength, oxygen availability, redox modifiers, and neuronal geometry.

Before we proceed, an important theoretic note that has deep impact on experimental reasoning is to be discussed, as follows. Electrodes measuring transmembrane potential (or voltage; like a microelectrode inside a cell and a reference electrode outside) are designed to measure a Nernstian (electrolytic, potentiometric) equilibrium potential. They do not directly measure capacitive currents. However, the signal they measure (like the action potential, AP or excitatory postsynaptic potential, EPSP) is supposedly generated by ionic (electrolytic) currents flowing through resistive ion channels. These currents can purportedly secondarily charge and discharge the membrane capacitor. But, in reality, the electrode's job is to report the resulting voltage change across that capacitor with minimal interference. The membrane capacitor is the load that the ionic currents charge, but the electrode measures the voltage across that load via thermodynamic (Nernst) equilibrium, not capacitive current directly. Electrode placement affects the kinetics and fidelity of the measurement, but not the steady-state thermodynamic voltage, which is why experimenters can reproduce the same key values (like resting potential or AP peak). Table 1 shows the comparison of various aspects of experimental neuronal electrophysiology. Clearly, the measurements' reproducibility (experimental data agreements) is strong for near-equilibrium read-outs such as the resting potential; the action-potential peak, by contrast, is an explicitly non-equilibrium, time-varying capacitive transient and should not be read as a Nernstian equilibrium.

Table 1: Foundations of experimental electrophysiology and theoretical assumptions

\begin{longtable}[]{@{}
  >{\raggedright\arraybackslash}p{(\columnwidth - 6\tabcolsep) * \real{0.2500}}
  >{\raggedright\arraybackslash}p{(\columnwidth - 6\tabcolsep) * \real{0.2500}}
  >{\raggedright\arraybackslash}p{(\columnwidth - 6\tabcolsep) * \real{0.2500}}
  >{\raggedright\arraybackslash}p{(\columnwidth - 6\tabcolsep) * \real{0.2500}}@{}}
\toprule\noalign{}
\begin{minipage}[b]{\linewidth}\raggedright
Feature
\end{minipage} & \begin{minipage}[b]{\linewidth}\raggedright
Is it Electrolytic?
\end{minipage} & \begin{minipage}[b]{\linewidth}\raggedright
Is it Capacitive?
\end{minipage} & \begin{minipage}[b]{\linewidth}\raggedright
Role of Electrode Placement
\end{minipage} \\
\midrule\noalign{}
\endhead
\bottomrule\noalign{}
\endlastfoot
Resting Potential & Yes (Nernst equilibrium of K⁺) & No (Steady state, no current) & Minimal (if properly compensated) \\
Action Potential Peak & Approaches \(E_{Na}\) (never reached); not a true equilibrium & Yes (non-equilibrium capacitive transient; large net current near the peak) & Minimal \\
Rate of Rise (dV/dt) & Primarily No; Secondarily Yes. (Determined by Na⁺ current magnitude) & Primarily Yes (Determined by C\textsubscript{m}: dV/dt = I/C\textsubscript{m}) & Major effect~(series resistance) \\
Extracellular Recording (EKG) & Yes (Resistive voltage drop in saline) & No (Volume conduction of ionic currents) & Huge effect~(distance, orientation) \\
\end{longtable}

\section*{2. Comparison of the new theory with earlier treatments}

A. Relevance across the scales/types: To start off with, the electricality of a neuron is not an isolated phenomenon and the action potential waveforms are noted even in manmade synthetic structures and in plants. Fox microspheres, reconstituted lipid-proteinoid microstructures that are devoid of ion-channels (Przybylski et al., 1982; Przybylski \& Fox, 1984) showed animal-like temporal-amplitude signature of action potential waveforms. When monitored, plant-systems that show tactile movements (like \emph{Dionaea muscipula} or the Venus flytrap and \emph{Mimosa pudica} or ``Touch-me-not'') and which lack purported ion-pumping proteins like Na/K ATPase (Sibaoka, 1991; Volkov et al., 2010; De Luccia, 2012; Hedrich \& Kreuzer, 2023) demonstrated action-potential traces. Furthermore, anaesthetics like diethyl ether and lidocaine were found to deleteriously affect the electrophysiology of plants' responses also (De Luccia, 2012). This is a strong physiology-based argument for a more fundamental mechanism common to all systems, and interfacial redox activity would therefore be a more appropriate explanation for the neuronal electricality phenomenon (Manoj et al., 2023; Manoj et al., 2026), rather than trans-membrane cation-pumping/channelling mechanisms.

\subsection*{B. Comparison with earlier mathematical treatments of chemico-physical phenomena}

The murburn master neuronal function/equation belongs mathematically to the reaction--diffusion class (like autocatalytic chemical waves or combustion fronts and Belousov--Zhabotinsky reactions), which explains why it naturally supports traveling waves, finite propagation speed, and robustness. The equivalents in the two systems are concentration, diffusion term, reaction term and chemical wave in reaction-diffusion model AND electron retaining capacity (EHP), ERP-mediated relay, EDP-ESP imbalance and action potential in the murburn master model. Unlike all equations compared in Table 2, the murburn field variable \emph{ϕ} represents a capacity or state variable (electron holding potential), not a transported quantity. This distinction eliminates mass-transport limitations (minuscule for electrons, compared to ions!), allowing neuronal conduction velocities of 10--200 m/s without invoking ion pumping or channel gating. Hodgkin--Huxley (HH, 1952) model gave the waveform phenomenology and NCV (by adopting the ``cable theory'' of Lord Kelvin). Although cable theory mathematically complements HH, cable theory extends HH mathematically, though a molecular mechanism for charge generation at the chemical level was not the primary focus of those frameworks. Murburn theory derives most features of neuronal electricality from a single equation. Further, in murburn, voltage is derived (it is not the driving force!), electrical activity is primarily electronic (although secondarily ionic!) and NCV emerges naturally from the same dynamical parameters governing spike generation. This is while in classical neuroscience, voltage is causal, electrical activity is primarily ionic and NCV is maintained with high-flux selective Na/K channels (Na inward and K outward; which in turn requires a highly intelligent and energy-expensive maintenance of Na/K differential across the membrane). This mechanism talso solicits implicitly coordinated gating mechanisms (requiring spatiotemporal opening/closing properties; and that too, based on a circular logic). In short, murburn replaces passive ion-fluxes with active redox electron activity and ion-fluxes are stochastic-thermodynamic events arising from the primary event.

\clearpage
\begin{center}
\rotatebox{90}{%
\begin{minipage}{\dimexpr\textheight-1cm\relax}
\footnotesize
\setlength{\tabcolsep}{4pt}

\emph{\textbf{Table 2: Comparative snapshot of the murburn master equation and related formulations in chemico-physics}} (SV = state variable; similarities and differences are listed wrt to the first case in the top-row, the murburn master equation)

\medskip

\begin{tabular}{@{}
  >{\raggedright\arraybackslash}p{(\linewidth - 14\tabcolsep) * \real{0.12}}
  >{\raggedright\arraybackslash}p{(\linewidth - 14\tabcolsep) * \real{0.18}}
  >{\raggedright\arraybackslash}p{(\linewidth - 14\tabcolsep) * \real{0.09}}
  >{\raggedright\arraybackslash}p{(\linewidth - 14\tabcolsep) * \real{0.045}}
  >{\raggedright\arraybackslash}p{(\linewidth - 14\tabcolsep) * \real{0.12}}
  >{\raggedright\arraybackslash}p{(\linewidth - 14\tabcolsep) * \real{0.10}}
  >{\raggedright\arraybackslash}p{(\linewidth - 14\tabcolsep) * \real{0.135}}
  >{\raggedright\arraybackslash}p{(\linewidth - 14\tabcolsep) * \real{0.14}}@{}}
\toprule
\begin{minipage}[b]{\linewidth}\raggedright
Model
\end{minipage} & \begin{minipage}[b]{\linewidth}\raggedright
Form
\end{minipage} & \begin{minipage}[b]{\linewidth}\raggedright
Context
\end{minipage} & \begin{minipage}[b]{\linewidth}\raggedright
SV
\end{minipage} & \begin{minipage}[b]{\linewidth}\raggedright
Meaning of Field
\end{minipage} & \begin{minipage}[b]{\linewidth}\raggedright
Propagation Mechanism
\end{minipage} & \begin{minipage}[b]{\linewidth}\raggedright
Similarity
\end{minipage} & \begin{minipage}[b]{\linewidth}\raggedright
Difference
\end{minipage} \\
\midrule
Murburn Master Equation (This work) & \emph{∂\textsubscript{t}\hspace{0pt}ϕ} = \(\mathcal{D}\text{ }\nabla^{'2}u + \Theta\left( \gamma + \alpha u - \beta^{'}u^{2} \right)\) & Neuronal electrophysiology (this work) & \emph{ϕ} & Electron Holding Potential (redox capacity) & Electronic redox wave & Unified waveform + velocity; traveling waves & Field is charge-holding capacity, not mass or voltage \\
Reaction--Diffusion Equation (Turing, 1952) & \emph{∂\textsubscript{t}\hspace{0pt}u} = \emph{D}∇\textsuperscript{2}\emph{u} + \emph{f(u)} & Chemical kinetics, pattern formation & \emph{u} & Concentration of reacting species & Molecular diffusion & Excitable waves, pattern formation & Requires mass transport; slow by diffusion limits \\
Fisher--KPP Equation (Fisher, 1937; Kolmogorov, 1937) & \emph{∂\textsubscript{t}\hspace{0pt}u} = \emph{D}∇\textsuperscript{2}\emph{u} + \emph{ru (1−u)} & Population dynamics, gene spread & \emph{u} & Population density & Growth + diffusion & Closed-form wave speed \emph{v} = 2(\emph{Dr})\textsuperscript{½} \hspace{0pt} & Logistic saturation has no redox analogue \\
Excitable Media Models (FitzHugh 1961, Nagumo et al., 1962) & \emph{∂\textsubscript{t}\hspace{0pt}u} = \emph{D}∇\textsuperscript{2}\emph{u} + \emph{g(u)} & Cardiac tissue, calcium waves & \emph{u} & Excitability variable & Reaction--diffusion & Threshold, refractoriness, wave annihilation & Excitability postulated, not chemically derived \\
Cable Equation (Thomson 1855, Rall, 1962) & \emph{∂\textsubscript{t}\hspace{0pt}V} = \emph{D}∇\textsuperscript{2}\emph{V} − \emph{V/τ} & Classical neuroscience (CMPT) & \emph{V} & Membrane voltage & Passive ionic current & Diffusion + relaxation structure & Voltage is causal, NCV too slow without channels \\
Ginzburg--Landau Equation (Ginzburg \& Landau, 1950) & \emph{∂\textsubscript{t}\hspace{0pt}ψ} = \emph{D}∇\textsuperscript{2}\emph{ψ} + \emph{aψ} − \emph{bψ\textsuperscript{3}} & Phase transitions, condensed matter & \emph{ψ} & Order parameter & Domain growth & Phase fronts, criticality & No chemical redox or biological meaning \\
Heat Equation (Fourier, 1822) & \emph{∂\textsubscript{t}\hspace{0pt}T} = \emph{κ}∇\textsuperscript{2}\emph{T} + \emph{S} & Thermal physics & \emph{T} & Temperature & Thermal diffusion & Diffusion + source term & No self-amplification or excitability \\
\bottomrule
\end{tabular}
\end{minipage}%
}
\end{center}
\clearpage

As seen from Table 2, the master murburn neuronal activity equation is no \emph{ad hoc} concoction, but is a derivation that lives in a deeply studied mathematical class of relations in nature, and inherits known results/features: traveling waves, finite propagation speeds, robustness to noise, scaling laws, etc. Yet it differs in physical interpretation, which is where the novelty lies for the murburn model. The new model presents mechanistically sound redox-defined excitability (which can also be applied in other excitable systems like cardiac and other muscles, and explain phenomena like `calcium waves'), electron-relay in heterogeneous aqueous phase (sans ion-flux causations; with voltage as a derived variable, not the driving force for any ionic-motilities!) and a single equation providing the major electricality aspects of neuron. We note that Table~2 places the bare single-field equation in the Fisher--KPP (monostable front) class; the two-variable excitable extension of Part~1 (Section~XI), combining a bistable reaction with a slow recovery variable, then moves the full model into the FitzHugh--Nagumo (excitable-media) row, the bistable reaction supplying the threshold and the recovery variable the self-resetting spike and refractory period. The taxonomy is thus internally consistent: the bare single field yields fronts, and the excitable extension yields the spike.

\subsection*{C. Comparison of murburn model with other/earlier frameworks for neuronal function}

Among the four major frameworks proposed for neuronal impulse conduction, the murburn model uniquely provides a chemically grounded, thermodynamically consistent, and quantitatively unified explanation of both the action-potential waveform and conduction velocity (Table 3). Unlike HH--cable, soliton, or electromagnetic models, murburn theory naturally resolves charge-counting discrepancies, reversible heat observations, redox sensitivity, and the modest magnitude of ionic fluxes within a single redox-electronic master equation.

\emph{\textbf{Table 3: A comparison of the four major theories mooted in the research community for neuronal function}}

\begin{longtable}[]{@{}
  >{\raggedright\arraybackslash}p{(\columnwidth - 10\tabcolsep) * \real{0.0612}}
  >{\raggedright\arraybackslash}p{(\columnwidth - 10\tabcolsep) * \real{0.1863}}
  >{\raggedright\arraybackslash}p{(\columnwidth - 10\tabcolsep) * \real{0.2059}}
  >{\raggedright\arraybackslash}p{(\columnwidth - 10\tabcolsep) * \real{0.1790}}
  >{\raggedright\arraybackslash}p{(\columnwidth - 10\tabcolsep) * \real{0.1948}}
  >{\raggedright\arraybackslash}p{(\columnwidth - 10\tabcolsep) * \real{0.1728}}@{}}
\toprule\noalign{}
\begin{minipage}[b]{\linewidth}\raggedright
No.
\end{minipage} & \begin{minipage}[b]{\linewidth}\raggedright
Criteria
\end{minipage} & \begin{minipage}[b]{\linewidth}\raggedright
HH--Cable
\end{minipage} & \begin{minipage}[b]{\linewidth}\raggedright
Soliton
\end{minipage} & \begin{minipage}[b]{\linewidth}\raggedright
Electromagnetic wave
\end{minipage} & \begin{minipage}[b]{\linewidth}\raggedright
Murburn
\end{minipage} \\
\midrule\noalign{}
\endhead
\bottomrule\noalign{}
\endlastfoot
1 & Primary Proponents & Hodgkin \& Huxley (1952) & Heimburg \& Jackson (2005) & Various (field-based models) (Pockett 2000; McFadden, 2002; Jacak \& Jacak, 2020) & KMM et al. \\
2 & Key Physical Players & Na⁺, K⁺ ions; voltage-gated channels; membrane capacitance & Lipid bilayer elasticity; density pulses & Electric \& magnetic fields; displacement currents & Electrons; DRS; redox microdomains; ions as buffers \\
3 & Primary State Variable & Membrane voltage (V\textsubscript{m}) & Membrane density / thickness & Electric field / EM wave & Electron Holding Potential (EHP, \(\phi\)) \\
4 & Mode of Propagation & Ionic currents + passive cable spread & Nonlinear mechanical soliton & EM field propagation along axon & Redox-electronic relay (reaction--diffusion field) \\
5 & Nature of the Signal & Ionic--electrical & Mechanical--thermodynamic & Electromagnetic & Redox-electronic (chemical--electrical) \\
6 & Signal transduction & Not explicitly addressed; somehow different neurons are geared to collect signals for relay & Stimulus provides mechanical energy sufficient to nucleate a nonlinear density pulse in the lipid bilayer & External stimulus induces local electric or electromagnetic field perturbations in or near the axon & Thermal, mechanical, photonic or chemical perturbation of local oxygen-superoxide redox equilibria to alter EHP at axon hill \\
7 & Threshold phenomena & Arises from nonlinear voltage-dependent gating kinetics of Na⁺ channels; largely parameter fitted & Depends on membrane elastic constants and proximity to lipid phase transition & Arises from field strength exceeding dielectric or conductive limits & Triggering depends on signal magnitude and neuron geometry \\
8 & Energy Source & Electrochemical ion gradients (ATP-maintained) & Stored elastic energy in membrane & Field energy (poorly specified biologically) & Continuous metabolic redox activity (ECS/DRS) \\
9 & Charge Carrier & Ions crossing membrane & None (density pulse) & Displacement current & Electrons (primary); ions secondary \\
10 & Speed of Transmission & 0.5--120 m/s (requires channels \& myelin) & Comparable to sound in membrane (problematic scaling) & Near-light speed (biologically unrealistic) & 1--200 m/s naturally (geometry \& redox-limited) \\
11 & Explanation of AP Waveform & Fitted via gating kinetics & Emerges from nonlinear elasticity & Not intrinsic & Emerges from EHP collapse \& recovery \\
12 & Heat Generation & Predicts net heat dissipation & Predicts near-adiabatic propagation & Not addressed & Low heat; redox-buffered, near-reversible \\
13 & Experimental Heat Observations & Inconsistent (reversible heat observed) & Consistent & Not predictive & Consistent \\
14 & Metabolic Coupling & Indirect (pumps restore gradients later) & Weak / unclear & Largely ignored & Direct and continuous (oxygen, redox, DRS) \\
15 & Sensitivity to Oxygen / Redox State & Weak until ATP failure & Weak & Not specified & Strong and early \\
16 & Role of Na⁺/K⁺ Fluxes & Causal drivers & Secondary consequences & Often ignored & Secondary buffering responses \\
17 & Ionic Strength Dependence & Monotonic or saturating & Weak & Not defined & Bell-shaped (predictive) \\
18 & Spatial Constraints & Requires continuous membrane \& channels & Requires intact lipid order & Requires waveguides & Works in bulk cytosol + membrane \\
19 & Myelination Effect & Electrical insulation \& capacitance reduction & Mechanical impedance change & Field confinement & ERP coherence enhancement \\
20 & Charge Accounting & Ion flux insufficient (known mismatch) & Avoids charge issue & Avoids chemistry & Resolved via electronic charge redistribution \\
21 & Unified Explanation of Waveform + NCV & No; Separate models & No NCV derivation & No waveform theory & Yes; Single master equation \\
22 & Thermodynamic Consistency & Partial & Strong (mechanical) & Weak & Strong (chemical + physical) \\
23 & Falsifiable Predictions & Limited (parameter-fitted) & Mechanical signatures & Hard to test & Multiple decisive tests \\
24 & Overall Status & Phenomenological success, outstanding mechanistic questions & Elegant but incomplete in several aspects & Speculative & Unified, mechanistic, testable, experimentally supported \\
\end{longtable}

The complexity of neurons does not agree with the idealized geometry and uniformity that the classical perceptions seek (nor do the structure/distribution and function correlations of protein pumps/channels). The classical theory assumes an ohmic current, ignoring binding/diffusion constraints. Further, coupling of metabolisms to energy is lacking, and it also fails to account for the fundamental stochasticity that underpins the ``scheme of events''. In the murburn framework, charge is provided by electronic redistribution within the chemico-electromagnetic matrix (CEM) by ECS-DRS mechanism temporarily holding and releasing electrons, water and ions providing screening and stabilization (not bulk charge), and thus, in this framework, charge displacement ≠ ion transport! Most of the required charge comes from electronic rearrangement, which is fast and energetically cheap. Na⁺ and K⁺ actually respond to the redox-generated electric field, buffer and stabilize charge asymmetries, and slowly adjust gradients over many spikes. Their flux magnitudes are therefore appropriately small. In murburn, ionic fluxes are expected to be small, metabolic cost per spike is low, and it naturally explains why spikes are fast. What looks like a ``problem'' for HH is actually a confirmation of murburn logic.

\section*{3. Sensory transduction in the murburn framework}

All sensory stimuli ultimately perturb the CEM, thereby altering the oxygen--superoxide equilibrium and the distribution of diffusible reactive species.

\subsection*{Stimuli transduction}

CMPT and murburn theory make qualitatively incompatible predictions about causality, scaling, sensitivity and failure modes. 

The experiments suggested below (and summarized in \textbf{Table 5}) do not rely on fitting waveforms.

They test control-parameter dependence, where the two theories diverge. Across ionic, redox, geometric, thermal, and protonic perturbations, murburn theory yields falsifiable predictions that classical pump--channel models cannot accommodate without auxiliary assumptions.

The equilibrium between molecular oxygen and superoxide radical:

\[\text{O}_{2} + e^{-} \rightleftharpoons \text{O}_{2}^{\text{•} -}(E^{\circ} \approx - 0.16\text{ V})
\]serves as the primary ``cellular antenna'' because:

1. Ubiquity: O₂ is present everywhere (\textasciitilde30 µM in tissues)

2. Sensitivity: The O₂/O₂•⁻ couple sits at \textasciitilde-160 mV, in the middle of the biological redox window (-400 to +800 mV)

3. Kinetic lability: Superoxide has µs-ms lifetime, long enough to sense, short enough to reset

4. Connectivity: O₂•⁻ connects to all major redox circuits (NADH/NAD⁺, GSH/GSSG, ascorbate, heme proteins); Every stimulus ultimately perturbs this equilibrium by altering either: O₂ availability (concentration, diffusion), electron availability (redox state), superoxide lifetime (scavenging, dismutation)

\subsection*{A. Thermal Stimuli (Heat/Cold)}

1. Direct effect on equilibrium constant:

\[K_{\text{eq}}(T) = \exp(-\Delta G^{\circ}/RT),\qquad \Delta G^{\circ} = -F E^{\circ},\ \ E^{\circ} = -0.16\ \mathrm{V}\]

For ΔT = 1°C (37→38°C), \(K_{\text{eq}}\) shifts by of order a few percent per °C (an order-of-magnitude estimate; the rigorous temperature slope is the van\textquotesingle t Hoff relation \mbox{\(d\ln K/dT = \Delta H^{\circ}/RT^{2}\)}), and the superoxide concentration changes proportionally.

2. Indirect effects:

Membrane fluidity changes: Alters O₂ permeability (ΔP ≈ 2\%/°C)

Enzyme kinetics: SOD activity Q₁₀ ≈ 2, catalase Q₁₀ ≈ 1.5

Diffusion coefficients: D\textsubscript{O₂} increases \textasciitilde3\%/°C

Net effect: Heating shifts O₂/O₂•⁻ equilibrium toward more O₂•⁻ (higher {[}O₂•⁻{]}/{[}O₂{]} ratio) by Increasing thermal energy for electron transfer, Increasing O₂ diffusion into redox-active sites, temporarily overwhelming antioxidant capacity.

Example: In thermoreceptor neurons, Basal {[}O₂•⁻{]} ≈ 0.1 nM at 37°C. At 39°C: {[}O₂•⁻{]} rises by of order ten percent (a couple-degree rise gives a few-percent change per °C), enough to cross the ERP network threshold. This crosses ERP network threshold → action potentials. Cold has opposite effect: reduces {[}O₂•⁻{]} below baseline.

\subsection*{B. Mechanical stimuli (pressure, stretch, vibration)}

1. Membrane deformation effects:

Mechanical stress alters: Electrical activity via stretch-activated channels → changes electric field in vicinity, Lipid packing → alters O₂ solubility and diffusion, Membrane curvature → concentrates/disperses redox proteins/molecules.

Quantitative example (Pacinian corpuscle):

Pressure deforms membrane by \textasciitilde5 nm

Local electric field changes by ΔE ≈ 10⁶ V/m

This alters EDP of nearby redox centers by:

ΔEDP = exp (-\emph{zeΔEd}/(2\emph{k\textsubscript{B}T})) is a change of order ten percent; altered EDP changes O₂•⁻ production rate

2. Cytoskeletal transduction:

Microtubules and actin filaments are piezoelectric and redox-active: Stress generates electric potentials (∼mV per 1\% strain), This electric field directly modulates EDP/ERP of bound redox centers, Neuroglobin bound to microtubules has its ERP modified by mechanical stress

3. Mitochondrial compression:

Mechanical pressure on mitochondria: Alters cristae structure → changes electron transport chain efficiency, Increases ROS production 2-5 fold with 10\% compression, Directly perturbs O₂/O₂•⁻ equilibrium

\subsection*{C. Chemical stimuli (neurotransmitters, pH, ions)}

1. Neurotransmitter receptor Activation:

Glutamate (NMDA receptors): Ca²⁺ influx → activates mitochondrial ROS production, Quantitative: {[}Ca²⁺{]} from 0.1 µM → 1 µM increases O₂•⁻ production 10-fold, Threshold: \textasciitilde0.5 µM Ca²⁺ triggers ERP wave propagation

Dopamine (autoxidation): \(\text{Dopamine} + \text{O}_{2} \rightarrow \text{Dopamine-o-quinone} + \text{O}_{2}^{\text{•} -}\)\\
Each dopamine oxidized produces 0.5-2 O₂•⁻ molecules, Dopamine vesicles contain ∼10⁴ molecules → significant O₂•⁻ formation upon release

2. pH changes:

Protons directly participate in O₂•⁻ dismutation: \(2\text{O}_{2}^{\text{•} -} + 2\text{H}^{+} \rightarrow \text{H}_{2}\text{O}_{2} + \text{O}_{2}\)\\
Lower pH (more H⁺) favors dismutation → lowers {[}O₂•⁻{]}, Higher pH (fewer H⁺) slows dismutation → increases {[}O₂•⁻{]}; ΔpH of 0.1 changes {[}O₂•⁻{]} by ∼15\%

3. Ionic modulation:

K⁺ efflux during repolarization: Increases extracellular {[}K⁺{]} from 4→12 mM, Alters membrane potential → changes electric field locally, Modulates O₂•⁻ stability (K⁺ stabilizes O₂•⁻ ion pair), Cl⁻ influx: Can quench radicals (Cl⁻ + •OH → Cl• + OH⁻)

\subsection*{D. Photonic stimuli (light and other suitable radiations)}

1. Direct photo-redox chemistry:

Photons with energy \textgreater{} O₂ singlet-triplet gap (0.98 eV, λ \textless{} 1270 nm) can:

Excite O₂ to its singlet state (¹O₂) via the well-documented photosensitised pathway, which can then drive downstream one-electron redox chemistry.

2. Rhodopsin-like sensors:

Retinal activation in photoreceptors directly leads to superoxide production. Modulates ERP of associated redox chains.

Quantitative (Rod cells): One photon isomerizes one rhodopsin, triggers amplification cascade, this crosses ERP threshold for bipolar cell signaling.

\subsection*{E. Electromagnetic fields (non-ionizing radiation)}

1. Direct electron spin effects: O₂•⁻ has unpaired electron → sensitive to magnetic fields via radical pair mechanism:

\(S + T_{0} \rightleftharpoons T_{+} + T_{-}\) (schematically \({}^{1}\mathrm{O_{2}} + {}^{3}\mathrm{O_{2}} \leftrightarrow {}^{2}\mathrm{O_{2}^{+}} + {}^{2}\mathrm{O_{2}^{-}}\))

Magnetic fields (∼1 mT - 1 T) alter spin state interconversion rates, changing:

O₂•⁻ recombination rates (k\textsubscript{rec}), O₂•⁻ lifetime (τ), Effective {[}O₂•⁻{]} by up to 50\% at 100 mT,

2. Induced electric fields: Time-varying B fields induce E fields (Faraday\textquotesingle s law):

\[\nabla \times \overrightarrow{E} = - \frac{\partial\overrightarrow{B}}{\partial t}
\]For 60 Hz, 1 mT: E induced ≈ 10⁻² V/m in cells; This modulates EDP/ESP of charged redox centers, Particularly affects neuroglobin ERP (heme is paramagnetic)

All stimuli converge on altering the chemico-electromagnetic-matrix (CEM), which then shifts the O₂/O₂•⁻ equilibrium, giving the general transduction equation:

\[\frac{d\lbrack\text{O}_{2}^{\text{•} -}\rbrack}{dt} = D\nabla^{2}\lbrack\text{O}_{2}^{\text{•} -}\rbrack + (k_{\text{prod}} + \Delta k_{\text{stim}}) - (k_{\text{dism}} + \Delta k_{\text{scav}})\lbrack\text{O}_{2}^{\text{•} -}\rbrack
\]Where stimulus effects appear as:

Δk\textsubscript{stim} = stimulus-modified production rate

Δk\textsubscript{scav} = stimulus-modified scavenging rate

\textbf{Table 4: Stimulus-specific Δk values}

\begin{longtable}[]{@{}
  >{\raggedright\arraybackslash}p{(\columnwidth - 6\tabcolsep) * \real{0.3297}}
  >{\raggedright\arraybackslash}p{(\columnwidth - 6\tabcolsep) * \real{0.1358}}
  >{\raggedright\arraybackslash}p{(\columnwidth - 6\tabcolsep) * \real{0.1503}}
  >{\raggedright\arraybackslash}p{(\columnwidth - 6\tabcolsep) * \real{0.3842}}@{}}
\toprule\noalign{}
\begin{minipage}[b]{\linewidth}\raggedright
\emph{Stimulus}
\end{minipage} & \begin{minipage}[b]{\linewidth}\raggedright
\emph{Δk\_prod (s⁻¹)}
\end{minipage} & \begin{minipage}[b]{\linewidth}\raggedright
\emph{Δk\_scav (M⁻¹s⁻¹)}
\end{minipage} & \begin{minipage}[b]{\linewidth}\raggedright
\emph{Mechanism}
\end{minipage} \\
\midrule\noalign{}
\endhead
\bottomrule\noalign{}
\endlastfoot
Heat (ΔT=+3°C) & +3×10⁻³ & -1×10⁶ & Increased ET kinetics, decreased SOD efficiency \\
Pressure (1 kPa) & +1×10⁻² & +5×10⁵ & Membrane deformation → increased ET complex proximity \\
Glutamate (1 mM) & +5×10⁻² & -2×10⁶ & Ca²⁺ influx → mitochondrial ROS burst \\
Light radiation (500 nm, 10¹⁴ photons/cm²/s) & +1×10⁻¹ & 0 & Direct photo-redox \\
Magnetic field (100 mT) & 0 & -1×10⁷ & Radical pair effect on recombination \\
\end{longtable}

\section*{4. Emergence of all-or-none behavior and yet, variable sensitivity}

The O₂/O₂•⁻ equilibrium provides~thresholding~via its connection to the~ERP network, where the critical ERP Threshold condition can be taken as-

\[\frac{\left\lbrack \text{O}_{2}^{\text{•} -}\rbrack_{\text{local}} \right.\ }{\left\lbrack \text{O}_{2}\rbrack_{\text{local}} \right.\ } > \left( \frac{\left\lbrack \text{O}_{2}^{\text{•} -} \right\rbrack}{\left\lbrack \text{O}_{2} \right\rbrack} \right)_{\text{crit}} \approx 10^{- 8}
\]

Neuron-type specific thresholds arise from:

1. Antioxidant capacity differences:

\[\text{Threshold} \propto \frac{\left\lbrack \text{SOD}\rbrack + \lbrack\text{Catalase}\rbrack + \lbrack\text{Peroxiredoxin} \right\rbrack}{\left\lbrack \text{ECS Agents} \right\rbrack}
\]

\begin{itemize}
\item
  Nociceptors (pain):~Low antioxidants → high sensitivity (threshold ≈ 10⁻⁹ M O₂•⁻)
\item
  Proprioceptors:~High antioxidants → low sensitivity (threshold ≈ 10⁻⁷ M O₂•⁻)
\item
  Photoreceptors:~Very high antioxidants → need 10⁶ photons to reach threshold
\end{itemize}

2. Structural differences:

\begin{itemize}
\item
  Myelination level:~Myelin contains antioxidants (vitamin E) → raises threshold
\item
  Mitochondrial density:~More mitochondria = more basal O₂•⁻ → lower threshold
\item
  Neuroglobin concentration:~High {[}Ngb{]} increases ERP efficiency → lowers threshold
\end{itemize}

3. Metabolic state differences:

\begin{itemize}
\item
  High ATP/ADP ratio:~Inhibits mitochondrial ROS → raises threshold
\item
  Low NADH/NAD⁺:~Reduces EDP sources → raises threshold
\end{itemize}

The oxygen-water cycle as the logic gate functions as a four-stage redox logic gate.

\[\text{O}_{2}\overset{\phantom{e^{-}}}{\rightarrow}\text{O}_{2}^{\text{•} -}\overset{\phantom{e^{-}}}{\rightarrow}\text{H}_{2}\text{O}_{2}\overset{\phantom{e^{-}}}{\rightarrow}\text{•OH}\overset{\phantom{e^{-}}}{\rightarrow}\text{H}_{2}\text{O}
\]\emph{i.e.}:

\textbf{1.} O₂ + e- → O₂•⁻\\
\textbf{2.} O₂•⁻+ 2H\textsuperscript{+} + e- → H₂O₂\\
\textbf{3.} H₂O₂ + H\textsuperscript{+} + e- → •OH + H₂O\\
\textbf{4.} •OH + H\textsuperscript{+} + e- → H₂O

Stage-dependent effects:

\begin{enumerate}
\def\labelenumi{\arabic{enumi}.}
\item
  O₂ → O₂•⁻:~Sensing stage (most sensitive to stimuli)
\item
  O₂•⁻ → H₂O₂:~Amplification stage (SOD catalyzes, can be rate-limiting)
\item
  H₂O₂ → •OH:~Decision stage (Fenton chemistry, irreversible commitment)
\item
  •OH → H₂O:~Execution stage (damage or signaling execution)
\end{enumerate}

Different stimuli create~characteristic redox waveforms:

1. Brief mechanical stimulus: Sharp {[}O₂•⁻{]} spike (ms duration), little H₂O₂ production, and fast ERP wave propagation

2. Sustained chemical stimulus: Sustained {[}O₂•⁻{]} elevation (seconds), significant H₂O₂ buildup, slower-broader ERP wave

3. Thermal stimulus: Gradual {[}O₂•⁻{]} rise (seconds-minutes), proportional H₂O₂ increase, Temperature-dependent ERP velocity

Sample predictions of the above model considerations:

1. All sensory neurons should show stimulus-induced ROS production before ion fluxes.

2. Antioxidants should raise sensory thresholds proportionally to their O₂•⁻ scavenging rates.

3. Magnetic fields should modulate sensory sensitivity in a frequency-dependent manner.

4. Temperature sensitivity should correlate with mitochondrial density across neuron types.

5. Deuterium oxide should alter sensory thresholds (proton-coupled ET effects).

Evidence in literature to support the above predictions:

Thermoreception:~TRPV channels are redox-sensitive; antioxidants block thermal responses.

Mechanoreception:~Stretch increases mitochondrial ROS within 10 ms.

Photoreception:~Retinal ROS production precedes electrical responses.

Chemoreception:~Odorant receptors trigger ROS bursts in olfactory neurons.

Nociception:~Pain stimuli increase O₂•⁻; SOD inhibitors increase pain sensitivity.

\section*{5. Validation/falsification predictions and comparative analysis}

The simplest of the distinctions is to isolate the primary and faster electronic event from the secondary ionic movement. In neurons, the observations of action potentials are redox waves with electron flow as primary signal, and minute ion movements are compensatory flows at surfaces for maintaining electroneutrality, resulting from hydration shell changes. That is, the apparent ``ionic conduction'' is actually a macroscopic approximation of nano-micro scale of stochastic electron networks. It will take many orders more time for intact ions to hop from one node to another in saltatory progression. Therefore, if you could measure electron flow and ion flow separately in a neuron during an action potential: (1) Electron flow would precede ion flow by \textasciitilde1 µs; Ion flow velocity would be diffusion-limited (\textasciitilde mm/s) \& (2) Electron flow velocity would match action potential velocity (\textasciitilde50-100 m/s); Blocking ion channels would alter AP shape but not prevent redox wave propagation.

Further, the origin and divergence of HH and murburn models for simple experimental conditions enable their separation owing to predictive powers, as delineated in Table 5. (Supplementary Information gives details of expected outcomes and also provides Python-based simulation outputs for the action potential waveform (Figure S1), spatial propagation dynamics (Figures S2--S3), and the predicted impact of five experimental parameters on NCV (Figure S4), as per the murburn model)

\emph{\textbf{Table 5: Predictions and impacts of HH and murburn models wrt various experimental variables}}

\begin{longtable}[]{@{}
  >{\raggedright\arraybackslash}p{(\columnwidth - 4\tabcolsep) * \real{0.3333}}
  >{\raggedright\arraybackslash}p{(\columnwidth - 4\tabcolsep) * \real{0.3333}}
  >{\raggedright\arraybackslash}p{(\columnwidth - 4\tabcolsep) * \real{0.3334}}@{}}
\toprule\noalign{}
\begin{minipage}[b]{\linewidth}\raggedright
\textbf{Perturbation}
\end{minipage} & \begin{minipage}[b]{\linewidth}\raggedright
\textbf{HH prediction}
\end{minipage} & \begin{minipage}[b]{\linewidth}\raggedright
\textbf{murburn prediction}
\end{minipage} \\
\midrule\noalign{}
\endhead
\bottomrule\noalign{}
\endlastfoot
ROS scavengers & little effect & reduced excitability \\
oxygen depletion & late effect & immediate excitability loss \\
D₂O solvent & little change & altered spike kinetics \\
ionic strength & monotonic effect & bell-shaped NCV \\
\end{longtable}

As per the murburn model, the thermodynamic/statistical (not protein-selective) inward movement of Na+ during the depolarization and the thermodynamic/statistical (not protein-selective) outward movement of K\textsuperscript{+} during the re-/hyper-polarization is not a voltage-dependent activity of the membrane-channel, but a concerted surface-capacitive behaviour of the bulk interior aqueous phase which transmits electrons through, as seen from Table 5. Experimentally, it should be verifiable that the flux of ions during various phases follows a statistical distribution, and not a deterministic selective channelling or pumping. These predictions are quantitatively supported by numerical simulations of the spatial Γ-relaxation PDE across five experimental parameters (Figure S4). The model reproduces: (A) monotonic increase in NCV with axon diameter, consistent with D ∝ d² scaling; (B) a bell-shaped NCV dependence on ionic strength, peaking near 125--150 mM, a prediction with no analogue in HH-cable theory; (C) monotonic NCV increase with temperature, consistent with Q₁₀ ≈ 2.5 (Table 5); (D) increasing NCV with internodal distance under myelination; and (E) exponential sensitivity of NCV to the driving redox potential E₀.

\emph{\textbf{Table 6: Murburn explanations for the surface movement of ions during various stages of action potential.}}

\begin{longtable}[]{@{}
  >{\raggedright\arraybackslash}p{(\columnwidth - 6\tabcolsep) * \real{0.2500}}
  >{\raggedright\arraybackslash}p{(\columnwidth - 6\tabcolsep) * \real{0.2500}}
  >{\raggedright\arraybackslash}p{(\columnwidth - 6\tabcolsep) * \real{0.2500}}
  >{\raggedright\arraybackslash}p{(\columnwidth - 6\tabcolsep) * \real{0.2500}}@{}}
\toprule\noalign{}
\begin{minipage}[b]{\linewidth}\raggedright
Phase
\end{minipage} & \begin{minipage}[b]{\linewidth}\raggedright
Redox state
\end{minipage} & \begin{minipage}[b]{\linewidth}\raggedright
Role of Na⁺
\end{minipage} & \begin{minipage}[b]{\linewidth}\raggedright
Role of K⁺
\end{minipage} \\
\midrule\noalign{}
\endhead
\bottomrule\noalign{}
\endlastfoot
Rest & Steady-state balanced EHP & Distributed extracellular buffer & Intracellular charge stabilizer \\
Depolarization & EHP collapse (EDP) & Rapid electrostatic compensation & Largely retained \\
Peak & Minimal EHP & Weak stabilization & Minimal participation \\
Repolarization & ESP dominance & Withdraws & Begins charge discharge \\
Hyperpolarization & EHP overshoot & Minimal role & Strong stabilizer, partial efflux \\
Recovery & Steady-state redox balance restoration & Gradients reset & Structural role resumes \\
\end{longtable}

Tables 6-8 below compare the various criteria with respect to the classical and murburn purviews.

\textbf{Table 7: Comparing the essence of mathematical formulation}

{\small
\begin{longtable}[]{@{}
  >{\raggedright\arraybackslash}p{(\columnwidth - 6\tabcolsep) * \real{0.2295}}
  >{\raggedright\arraybackslash}p{(\columnwidth - 6\tabcolsep) * \real{0.2953}}
  >{\raggedright\arraybackslash}p{(\columnwidth - 6\tabcolsep) * \real{0.2155}}
  >{\raggedright\arraybackslash}p{(\columnwidth - 6\tabcolsep) * \real{0.2598}}@{}}
\toprule\noalign{}
\begin{minipage}[b]{\linewidth}\raggedright
\textbf{Feature}
\end{minipage} & \begin{minipage}[b]{\linewidth}\raggedright
\textbf{HH--Cable (List \& No.)}
\end{minipage} & \begin{minipage}[b]{\linewidth}\raggedright
\textbf{Murburn (List \& No.)}
\end{minipage} & \begin{minipage}[b]{\linewidth}\raggedright
\textbf{Interpretive Comments}
\end{minipage} \\
\midrule\noalign{}
\endhead
\bottomrule\noalign{}
\endlastfoot
\textbf{Field variable} & \(V(x,t)\)membrane voltage (1) & \(u(x,t)\) or \(\phi(x,t)\ \)redox/EHP field (1) & HH uses voltage as the primary dynamical field; murburn uses a dimensionless thermodynamic/redox field logarithmically related to voltage. \\
\textbf{Other dynamic variables} & \(m,h,n;\ \)gating variables (3) & None in minimal formulation (0) & HH requires auxiliary channel-state variables; murburn embeds amplification and saturation intrinsically in the nonlinear reaction term. \\
\textbf{Total dynamic variables} & \(V,m,h,n\ \)(4 total) & \(u\ \)only (1 total) & HH is a coupled multi-variable dynamical system; murburn is a single-field nonlinear PDE. \\
\textbf{Spatial variable(s)} & \(x\ \)(1) & \(x\ \)(1) & Both are spatially distributed propagation models. \\
\textbf{Temporal variable(s)} & \(t\) (1) & \(t\ \)(1) & Both are time-evolution systems. \\
\textbf{Traveling-wave variable} & Usually implicit or separately derived (0--1) & \(U(z)\), \(z = x - vt\ \)(2 auxiliary variables) & Traveling-wave structure is intrinsic and explicit in murburn formulation. \\
\textbf{Principal parameters} & \({\overset{ˉ}{g}}_{Na},{\overset{ˉ}{g}}_{K},g_{L},C_{m},R_{i},a,I_{ext}\ \)(\textasciitilde7 core parameters) & \(D,\tau,\alpha,\beta^{'},\gamma\ \)(5 core parameters) & HH parameters are conductance/capacitance based; murburn parameters represent transport, kinetics, amplification, saturation, and basal drive. \\
\textbf{Voltage-dependent rate parameters} & \(\alpha_{m},\beta_{m},\alpha_{h},\beta_{h},\alpha_{n},\beta_{n}\)with multiple fitted coefficients (\textasciitilde12--18 effective parameters) & None (0) & HH relies heavily on empirically fitted voltage-rate relations; murburn avoids explicit gating-rate functions. \\
\textbf{Equilibrium/reversal constants} & \(E_{Na},E_{K},E_{L}\ \)(3) & \(\phi_{ref}\), \(RT/F\ \)(2) & HH uses ion-specific reversal potentials; murburn uses a thermodynamic scaling relation. \\
\textbf{Characteristic scales/constants} & \(C_{m},a,R_{i\ }\)(3) & \(L,\tau_{0}\ \)(2) & HH uses cable-theoretic electrical constants; murburn uses scaling parameters for nondimensionalization. \\
\textbf{Nonlinear terms} & \(m^{3}h(V - E_{Na}),n^{4}(V - E_{K})\) & \(\alpha u - \beta^{'}u^{2}\) & HH nonlinearity emerges through gating kinetics; murburn uses explicit local nonlinear feedback and saturation. \\
\textbf{Reaction term} & Ionic current balance (implicit) & Explicit: \(f(u) = \gamma + \alpha u - \beta^{'}u^{2}\) & Murburn directly separates transport and local reaction dynamics. \\
\textbf{Transport term} & Cable-current diffusion term & Reaction--transport diffusion term \(D\nabla^{2}u\) & HH transport represents ionic current flow; murburn transport represents effective redox/electron-relay coupling. \\
\textbf{Relaxation term} & Implicit through conductance dynamics & Explicit through \(\tau\) & Murburn contains an explicit relaxation timescale. \\
\textbf{Threshold mechanism} & Emergent from gating interplay & Explicit nonlinear instability criterion \(f(u) > 0\) & Thresholding is mathematically more transparent in murburn formalism. \\
\textbf{Resting-state condition} & Net ionic current balance & Stable fixed point \(f(u_{rest}) = 0\) & HH resting state depends on ion conductance equilibrium; murburn resting state is a nonlinear fixed point. \\
\textbf{Excitability criterion} & Voltage-dependent gating activation & Sign and slope structure of \(f(u)\) & HH excitability is channel-centric; murburn excitability is reaction-kinetic. \\
\textbf{Propagation condition} & Cable conductance + regenerative currents & Balance of amplification and transport & Murburn directly links NCV to transport and local growth. \\
\textbf{Closed-form NCV expression} & Approximate/derived numerically & \(v' = 2\sqrt{(D/\tau)\sigma}\) & Murburn yields an explicit analytical velocity relation. \\
\textbf{Waveform determinants} & Channel kinetics + conductance ratios & \(\tau,\alpha,\beta^{'},\gamma\) & Murburn maps waveform morphology directly to a compact parameter set. \\
\textbf{Dependence on ion specificity} & Explicit and central & Not explicit & HH fundamentally depends on Na/K formalism; murburn is chemically more generalized. \\
\textbf{Dependence on membrane permeability assumptions} & Central & Absent & HH depends on permeability/conductance concepts; murburn is bulk/membrane-based redox-relaxation dynamics. \\
\textbf{Metabolic/redox coupling} & Indirect/minimal & Direct and intrinsic & Murburn explicitly links metabolism/redox state to electrophysiology. \\
\textbf{Total explicit core entities} & \textasciitilde4 variables + \textasciitilde20--30 parameters/constants & \textasciitilde1 variable + \textasciitilde5--6 parameters/constants & Murburn provides a substantially lower-dimensional formulation. \\
\textbf{Core physical interpretation} & Neuron as an ion-conducting electrical cable & Neuron as a chemically driven excitable redox medium & The two frameworks differ fundamentally in mechanistic interpretation. \\
\end{longtable}
}

\textbf{Table 8: Comparison of two models and murburn advantages}

\begin{longtable}[]{@{}
  >{\raggedright\arraybackslash}p{(\columnwidth - 8\tabcolsep) * \real{0.0599}}
  >{\raggedright\arraybackslash}p{(\columnwidth - 8\tabcolsep) * \real{0.2178}}
  >{\raggedright\arraybackslash}p{(\columnwidth - 8\tabcolsep) * \real{0.2223}}
  >{\raggedright\arraybackslash}p{(\columnwidth - 8\tabcolsep) * \real{0.2579}}
  >{\raggedright\arraybackslash}p{(\columnwidth - 8\tabcolsep) * \real{0.2421}}@{}}
\toprule\noalign{}
\begin{minipage}[b]{\linewidth}\raggedright
No.
\end{minipage} & \begin{minipage}[b]{\linewidth}\raggedright
Criterion / Phenomenon
\end{minipage} & \begin{minipage}[b]{\linewidth}\raggedright
Classical CMPT (GHK--HH--Cable) Explanation
\end{minipage} & \begin{minipage}[b]{\linewidth}\raggedright
Murburn Redox Model Explanation / Prediction
\end{minipage} & \begin{minipage}[b]{\linewidth}\raggedright
Interpretation / Advantage in Murburn
\end{minipage} \\
\midrule\noalign{}
\endhead
\bottomrule\noalign{}
\endlastfoot
1 & \textbf{Origin of resting TMP} & Ionic concentration gradients + selective permeability of K⁺, Na⁺, Cl⁻ across membrane & Effective charge separation (ECS) driven by interfacial redox reactions involving oxygen and diffusible reactive species (DRS) & TMP emerges from \textbf{electron transfer chemistry}, not only ion gradients \\
2 & \textbf{Source of charge separation} & Transmembrane ionic redistribution & Electron transfer across membrane interface and DROS-mediated charge displacement & Explains potentials without requiring large ion movements \\
3 & \textbf{Maintenance of ionic gradients} & Na⁺/K⁺ ATPase pumping ions against gradients & Redox-mediated ion-pair solubility, protein adsorption, and metabolic equilibria determine ion distributions & Avoids need for high-energy stoichiometric pumping \\
4 & \textbf{K⁺\textgreater Na⁺ intracellular distribution} & Result of pump activity and selective channels & Combination of protein adsorption, phosphate pairing, and redox equilibria favors K⁺ retention & Explains distribution thermodynamically rather than mechanistically \\
5 & \textbf{Ion distribution order (K\textgreater Na\textgreater Mg\textgreater Ca)} & Not predicted from first principles & Predicted from ion-pair solubility and metabolic anion interactions & Provides \textbf{chemical basis} for ion hierarchy \\
6 & \textbf{Action potential depolarization} & Voltage-gated Na⁺ channels open → Na influx & Local redox perturbation changes DROS balance causing interfacial charge redistribution & Depolarization can arise without obligatory Na entry \\
7 & \textbf{Repolarization} & K⁺ channel activation causes K⁺ efflux & Re-equilibration of redox species and electron transfer cycles & Voltage recovery tied to metabolic state \\
8 & \textbf{Triggering of spike} & Membrane voltage threshold activates channels & Threshold concentration of reactive species triggers redox cascade & Explains stimulus thresholds chemically \\
9 & \textbf{All-or-none response} & Nonlinear gating of voltage-dependent channels & Nonlinear redox reaction thresholds and radical equilibria & Threshold chemistry naturally produces digital response \\
10 & \textbf{Variable sensitivity} & Variation in channel density and membrane properties & Variation in local redox environment, oxygen availability, ROS buffering & Explains metabolic modulation of excitability \\
11 & \textbf{Stimulus transduction (mechanical)} & Mechanical deformation opens mechanosensitive channels & Mechanical perturbation alters electron-transfer equilibria at membrane interface & Provides universal stimulus coupling mechanism \\
12 & \textbf{Thermal stimuli sensing} & Temperature-dependent channel kinetics & Temperature alters radical lifetimes and redox reaction rates & Predicts strong temperature sensitivity \\
13 & \textbf{Photonic stimulation} & Opsin-triggered G-protein cascades & Photon absorption perturbs redox equilibrium of chromophores and DROS & Direct photochemical explanation \\
14 & \textbf{Chemical stimuli detection} & Ligand binding to receptors & Molecules modulate DROS equilibrium and redox potential & Chemical sensing linked to redox perturbation \\
15 & \textbf{Electromagnetic stimuli} & Electric fields influence membrane potential & Fields perturb electron transfer equilibria and radical dynamics & Consistent with electromagnetic sensitivity \\
16 & \textbf{Axonal signal propagation} & Local circuit ionic currents along RC cable & Electron transfer cascade along membrane interface & Explains fast conduction without long ion diffusion \\
17 & \textbf{Conduction velocity limits} & Determined by membrane capacitance and axial resistance & Determined by redox relay rate and interface organization & Allows higher propagation efficiency \\
18 & \textbf{Myelination effects} & Reduces capacitance and leakage current & Restricts spatial degrees of freedom of redox relay & Both models predict faster conduction but murburn provides molecular rationale \\
19 & \textbf{Energy consumption during spikes} & Ionic pumping required to restore gradients & Minimal ionic redistribution; redox equilibration restores state & Lower metabolic cost \\
20 & \textbf{Spike propagation distance} & Requires repeated ionic currents & Electron relay along membrane interface & Explains long axonal conduction without large ion transport \\
21 & \textbf{Effect of oxygen availability} & Minor direct role in electrophysiology & Oxygen critical for DROS formation driving TMP and spikes & Predicts hypoxia strongly affects excitability \\
22 & \textbf{Effect of redox modulators} & Limited influence unless affecting channels & Radical scavengers or oxidants directly alter neuronal firing & Predicts strong pharmacological sensitivity \\
23 & \textbf{Effect of ROS scavengers} & Typically considered damaging or secondary & Direct regulators of electrophysiological behavior & Links oxidative stress and neuronal activity \\
24 & \textbf{pH dependence} & Mainly affects channel gating & Alters proton-coupled redox equilibria & Predicts stronger pH dependence \\
25 & \textbf{Solvent isotope effects (D₂O)} & Weak effects via viscosity & Alters radical chemistry and proton transfer & Predicts measurable electrophysiological changes \\
26 & \textbf{Ionic strength variation} & Affects conductance and screening & Alters redox potentials and ion pairing equilibria & Explains non-classical ionic effects \\
27 & \textbf{Effects of metabolic inhibitors} & Indirect via ATP depletion & Directly suppress redox-driven charge separation & Predicts immediate electrical changes \\
28 & \textbf{Effect of mitochondrial uncouplers} & Mainly metabolic effects & Direct perturbation of redox environment & Explains historical observations with DNP \\
29 & \textbf{Artificial systems showing spikes} & Difficult to reconcile & Redox interfaces can produce spike-like signals & Explains excitability in nonliving systems \\
30 & \textbf{Proteinoid / gel excitability} & Considered anomalies & Expected from redox-electronic charge separation & Supports universality of mechanism \\
31 & \textbf{TMP fluctuations in simple systems} & Not predicted & Redox reactions across interface generate potentials & Predicts electrophysiology in minimal systems \\
32 & \textbf{Coupling between metabolism and electrophysiology} & Indirect via ATP supply & Direct because redox metabolism drives potentials & Integrates bioenergetics and signaling \\
33 & \textbf{Evolutionary plausibility} & Requires complex pumps and channels & Requires simple redox chemistry and membranes & Simpler origin of excitability \\
34 & \textbf{Experimental predictions} & Channel blockade or ion substitution tests & DROS modulation, oxygen dependence, redox perturbation tests & Provides new experimental avenues \\
\end{longtable}

\textbf{Table 9: Dispositions of the two models}

\begin{longtable}[]{@{}
  >{\raggedright\arraybackslash}p{(\columnwidth - 6\tabcolsep) * \real{0.0484}}
  >{\raggedright\arraybackslash}p{(\columnwidth - 6\tabcolsep) * \real{0.3545}}
  >{\raggedright\arraybackslash}p{(\columnwidth - 6\tabcolsep) * \real{0.2893}}
  >{\raggedright\arraybackslash}p{(\columnwidth - 6\tabcolsep) * \real{0.3079}}@{}}
\toprule\noalign{}
\begin{minipage}[b]{\linewidth}\raggedright
\#
\end{minipage} & \begin{minipage}[b]{\linewidth}\raggedright
Observation
\end{minipage} & \begin{minipage}[b]{\linewidth}\raggedright
Difficulty for HH / CMPT
\end{minipage} & \begin{minipage}[b]{\linewidth}\raggedright
Natural murburn position
\end{minipage} \\
\midrule\noalign{}
\endhead
\bottomrule\noalign{}
\endlastfoot
1 & \textbf{Very small ionic imbalance needed to generate TMP} & HH attributes voltage primarily to ionic flux despite extremely small charge separation required & Voltage arises from electronic/redox charge redistribution rather than bulk ionic movement \\
2 & \textbf{Measured Na⁺ flux during spikes is tiny compared to total ionic pool} & HH assumes ionic flux drives the spike & Ionic flux is secondary electrostatic compensation \\
3 & \textbf{Reversible heat production during action potentials} & HH predicts net heat dissipation from ionic currents & Redox processes can be near-reversible with minimal heat \\
4 & \textbf{Rapid signal propagation compared with ionic diffusion rates} & HH relies on local circuit ionic currents; diffusion alone too slow & Electron relay in redox networks enables rapid propagation \\
5 & \textbf{Action potentials in organisms lacking classical Na⁺/K⁺ channel systems} (plants, protists) & HH framework specialized for animal ion channels & Redox-based excitability can occur in diverse biological systems \\
6 & \textbf{Excitable electrical behavior in artificial membranes or proteinoid microspheres} & HH requires specialized channels & Redox interfaces can generate electrical oscillations \\
7 & \textbf{Calcium-based action potentials in certain neurons and cardiac cells} & Requires alternative channel explanations & Excitability arises from general redox perturbation rather than specific ions \\
8 & \textbf{Effects of reactive oxygen species on neuronal membrane potential} & Treated as pathological side effects & ROS are central regulators of electron holding potential \\
9 & \textbf{Sensitivity of neuronal activity to metabolic inhibitors} & Explained indirectly via ATP depletion & Redox perturbation immediately affects electrical activity \\
10 & \textbf{Influence of deuterium oxide on nerve conduction} & Attributed mainly to viscosity effects & Changes proton-coupled electron transfer and radical dynamics \\
11 & \textbf{Temperature dependence of conduction velocity} & Modeled via channel kinetics & Redox reaction rates and radical lifetimes scale strongly with temperature \\
12 & \textbf{Magnetic-field effects on neural systems} & No clear mechanism in ionic models & Radical-pair spin chemistry can modulate redox reactions \\
13 & \textbf{Mismatch between predicted and measured metabolic cost of spikes} & HH predicts high ATP consumption & Electronic charge redistribution requires much less energy \\
14 & \textbf{Electrophysiological changes under oxidative stress conditions} & Considered secondary pathology & Reflect altered redox balance affecting EHP \\
15 & \textbf{Spike-like electrical oscillations in non-biological chemical systems (BZ reactions)} & Not related to ionic channels & Reaction--diffusion redox waves analogous to murburn dynamics \\
\end{longtable}

\section*{6. Published literature/awareness supporting murburn model}

Ritchie \& Straub (1980) found that AP conduction in non-myelinated mammalian nerve fibres is associated with O₂ consumption that correlates with conductance velocity, suggesting the involvement of redox processes. Howarth et al. (2012) reported that the energy cost of AP propagation in the brain does not tally with ion-pumping accounts. Further, redox effects have been correlated with excitability by researchers (Takahashi \& Copenhagen, 1996; Kourennyi et al., 2004), and neuroglobin appears to be a significant candidate in the redox dynamics of neurons (Wakasugi et al., 2003; Brunori et al., 2005). In murburn purview, the cell is not a bag of ions but a stochastic redox circuit where electrons flow (intra- and inter- molecular ETs are permitted on subnano- to suprapico- second timescales; or 10\textsuperscript{9} to 10\textsuperscript{13} per second; and electrons can hop at nm distances) through somewhat loosely-structured media, and in this view, ions adjust (in millisecond ranges, explaining the refractory period) to maintain the electrochemical conditions for that flow to continue. Assuming a distribution of 10 µM neuroglobin (present at 1 to 100 micromolar ranges in neurons, higher concentrations in more active neurons!), the distance between two neuroglobins is \textasciitilde50 nm. Key locations for mitochondrial clustering include synapses, axon initial segment, nodes of Ranvier, perinuclear region (soma) and dendritic branches. This aligns perfectly with murburn model in which neuroglobins and mitochondria are like transformers and power-stations.

Magnetic field effects: As it would impact radical dynamics, theoretically it is predictable that we could see some effect (Rodgers, 2009; Hore, 2012; Zhang et al., 2017) and there is literature available that quotes this outcome (Rosen, 2003; Mclean et al., 2003; Ghodbane et al., 2013).

Effect of antioxidants: Antioxidants are known to affect AP (and they should not have direct immediate effects on ionic conduction).

The axon hillock/dendrite region is geared for signal integration and initiation, while the synaptic end is adapted for signal output and neurotransmitter release. Axonal polarity is maintained by a combination of directed transport, cytoskeletal orientation, molecular filtering at the AIS, and selective membrane composition, ensuring distinct regional specialization. Also, at the dendrite and axon hillock, several redox-active proteins (SOD, catalase, peroxiredoxin, thioredoxin, glutaredoxin, etc.) and metabolites (nitoctinamide, glutathioine, etc.) are more enriched or functionally significant compared to the synaptic end (which has a higher concentration of neurotransmitters and calcium ions), reflecting their distinct roles in signal integration, action potential initiation, and metabolic regulation.

Now, we highlight two elaborate contexts: Effect of temperature (A) and heavy-water (B).

\begin{enumerate}
\def\labelenumi{\Alph{enumi}.}
\item
  \textbf{Temperature Dependence (Q₁₀) analysis}
\end{enumerate}

The temperature coefficient Q₁₀ quantifies how much a biological rate increases with a 10°C temperature rise:

Q\textsubscript{10} = (R\textsubscript{2}/R\textsubscript{1}) \textsuperscript{10/(T2 - T1)} ; where R₁, R₂ are rates at temperatures T₁, T₂.

Example calculation from data (Hodgkin \& Katz, 1949):

At 5°C: AP velocity = 12.5 m/s

At 15°C: AP velocity = 25.0 m/s

\[Q_{10} = \left( \frac{25.0}{12.5} \right)^{\frac{10}{10}} = 2.0
\]More recent mammalian data (cat motor neurons):

At 27°C: 45 m/s

At 37°C: 110 m/s

\[Q_{10} = \left( \frac{110}{45} \right)^{\frac{10}{10}} = 2.44\]

If AP propagation were limited by ion diffusion:

Viscosity of water: η(37°C)/η(27°C) ≈ 0.75/0.85 ≈ 0.88

Diffusion coefficient D \textasciitilde{} T/η

D(37°C)/D(27°C) = (310/300) × (0.85/0.75) ≈ 1.17

Expected Q₁₀ for diffusion-limited processes ≈ 1.17

But measured Q₁₀ ≈ 2.5, which indicates an activation energy barrier of:

\[E_{a} = RT^{2}\frac{\ln Q_{10}}{10} \approx (8.314)(310)^{2}\frac{\ln{2.5}}{10} \approx 73\text{ kJ/mol}
\]

This E\textsubscript{a} value matches PCET values of 50 to 70 kJ/mol (whereas E\textsubscript{a} values for diffusion or simple velocity-based processes are between 10 to 20 kJ/mol).

The observed result agrees more with murburn model. This high \mbox{\(Q_{10}\)} and the \mbox{\(\mathrm{D_2O}\)} isotope effect are best read as consistent with a redox/PCET step rather than as refuting HH, whose voltage-gated kinetics are themselves strongly temperature-dependent.

\begin{enumerate}
\def\labelenumi{\Alph{enumi}.}
\setcounter{enumi}{1}
\item
  \textbf{Solvent proton contribution (KIE) analysis}
\end{enumerate}

D₂O (heavy water)~differs from H₂O in two key ways relevant to nerve conduction:

\textbf{Higher viscosity:}~At 20°C, D₂O is about~1.25 times more viscous~than H₂O. This affects~\emph{hydrodynamic processes}~(like ion diffusion through aqueous pores).

\textbf{Different hydrogen bonding:}~The D--O bond is stronger than H--O (zero-point energy difference). This affects~\emph{proton/charge transfer}~kinetics and~\emph{redox chemistry}.

\begin{quote}
\(V_{H}\)~= conduction velocity in H₂O

\(V_{D}\)~= conduction velocity in D₂O

\(\eta_{H},\eta_{D}\)~= viscosities of H₂O and D₂O

\(k_{H},k_{D}\)~= rate-limiting step rate constants (redox/electron transfer) in H₂O and D₂O
\end{quote}

A. If conduction velocity~\(V\)~is proportional to ion diffusion rates:

\[V \propto D \propto \frac{1}{\eta}
\]Therefore,

\[\frac{V_{D}}{V_{H}} \approx \frac{\eta_{H}}{\eta_{D}}
\]

At 20°C:~\(\eta_{D}/\eta_{H} \approx 1.25\), so:

\[\frac{V_{D}}{V_{H}} \approx \frac{1}{1.25} = 0.80
\]Prediction: \textasciitilde20\% slowdown in D₂O, purely from viscosity.

B. If conduction velocity~\(V\)~is proportional to the rate of PCET:

\[V \propto k
\]For a proton-coupled electron transfer (likely in lipid redox reactions), the kinetic isotope effect (KIE) for H vs. D can be:

\[\text{KIE} = \frac{k_{H}}{k_{D}} \approx 2\text{ to }10\ \text{(often 4–7 at room T)}
\]This KIE arises from quantum tunnelling of protons, which is exponentially sensitive to mass.

This \(\frac{V_{D}}{V_{H}} \approx \frac{k_{D}}{k_{H}} \approx \frac{1}{\text{KIE}} \approx 0.1\text{ to }0.5\)

\emph{\hfill\break
}\textbf{Prediction:} 50--90\% slowdown in D₂O, due to the primary kinetic isotope effect on the redox reaction.

\textbf{Table 10:} The tables below capture the key findings of A (no. 1 through 4) \& B (no. 5 through 8).

\begin{longtable}[]{@{}
  >{\raggedright\arraybackslash}p{(\columnwidth - 6\tabcolsep) * \real{0.0632}}
  >{\raggedright\arraybackslash}p{(\columnwidth - 6\tabcolsep) * \real{0.3168}}
  >{\raggedright\arraybackslash}p{(\columnwidth - 6\tabcolsep) * \real{0.3748}}
  >{\raggedright\arraybackslash}p{(\columnwidth - 6\tabcolsep) * \real{0.2452}}@{}}
\toprule\noalign{}
\begin{minipage}[b]{\linewidth}\raggedright
\textbf{No.}
\end{minipage} & \begin{minipage}[b]{\linewidth}\raggedright
\textbf{Physiological experimental probe on neuronal impulse}
\end{minipage} & \begin{minipage}[b]{\linewidth}\raggedright
\textbf{Key finding}
\end{minipage} & \begin{minipage}[b]{\linewidth}\raggedright
\textbf{Reference}
\end{minipage} \\
\midrule\noalign{}
\endhead
\bottomrule\noalign{}
\endlastfoot
1 & Effect of temperature in squid giant axon & Velocity increased from \textasciitilde12.5 m/s at 5°C to \textasciitilde25 m/s at 15°C (Q₁₀ = 2.0) & Hodgkin \& Katz, 1949 \\
2 & Effect of temperature in cat's motor nerves & Velocity increased from \textasciitilde45 m/s at 27°C to \textasciitilde110 m/s at 37°C (Q₁₀ 2.3 to 3.0) & Paintal, 1965 \\
3 & Effect of temperature on myelinated nerves of \emph{Xenopus} & Q\textsubscript{10} for Na/K fluxes show values of 2.8 and 3.0 respectively, & Frankenhaeuser \& Moore, 1963 \\
4 & Effect of temperature on the nerves of cat's cornea & Q\textsubscript{10} for the nerves ranged from 2.3 to 2.8. & Chapman, 1967 \\
5 & Effect of D\textsubscript{2}O on frog nerves & Conduction velocity decreased by factor of \textasciitilde2.0 \& refractory period increased. & Spyropoulos, 1957 \\
6 & Effect of D\textsubscript{2}O on lobster nerve function & Conduction velocity decreased by factor of 1.8 to 2.2. & Moore, 1969. \\
7 & Effect of D\textsubscript{2}O on rat system & Conduction reduced by a factor 2 to 3. & BĂNg \& Mellander, 1976. \\
8 & Effect of D\textsubscript{2}O on squid giant axon & Conduction reduced by a factor 1.8 to 2.5. & Conti, 1970. \\
\end{longtable}

\textbf{Table 11: Summary of observations supports the murburn ET (proton-coupled), rather than ionic conduction!}

\begin{longtable}[]{@{}
  >{\raggedright\arraybackslash}p{(\columnwidth - 8\tabcolsep) * \real{0.2468}}
  >{\raggedright\arraybackslash}p{(\columnwidth - 8\tabcolsep) * \real{0.1859}}
  >{\raggedright\arraybackslash}p{(\columnwidth - 8\tabcolsep) * \real{0.2045}}
  >{\raggedright\arraybackslash}p{(\columnwidth - 8\tabcolsep) * \real{0.2315}}
  >{\raggedright\arraybackslash}p{(\columnwidth - 8\tabcolsep) * \real{0.1312}}@{}}
\toprule\noalign{}
\begin{minipage}[b]{\linewidth}\raggedright
\textbf{Parameter}
\end{minipage} & \begin{minipage}[b]{\linewidth}\raggedright
\textbf{Ion-Diffusion Prediction}
\end{minipage} & \begin{minipage}[b]{\linewidth}\raggedright
\textbf{Electron-Transfer Prediction}
\end{minipage} & \begin{minipage}[b]{\linewidth}\raggedright
\textbf{Experimental Result}
\end{minipage} & \begin{minipage}[b]{\linewidth}\raggedright
\textbf{Inference}
\end{minipage} \\
\midrule\noalign{}
\endhead
\bottomrule\noalign{}
\endlastfoot
\emph{\textbf{Effect of cooling by 10°C on charge transfer}} & 8\% slower (Q₁₀=1.1) & 60\% slower (Q₁₀=2.5) & 50-60\% slower (Q₁₀=2.3-2.8) & ET-theory appeals! \\
\emph{\textbf{Effect of solvent D₂O on mass transfer}} & 25\% slower (factor 0.75) & 67-80\% slower (factor 0.2-0.33) & 50-67\% slower (factor 0.33-0.5) & ET-theory appeals! \\
\end{longtable}

\section*{7. Physiological and technological implications}

The new theory is expected to allow us to develop better methodologies for pain management, sensory modulation and redox prosthetics. For example:

Redox-based sensory modification:

\begin{enumerate}
\def\labelenumi{\arabic{enumi}.}
\item
  Pain management:~Local antioxidants should increase pain thresholds.
\item
  Thermoregulation:~Mitochondrial uncouplers could modulate temperature sensitivity.
\item
  Photoreceptor protection:~Targeted O₂•⁻ scavengers for retinal diseases.
\item
  Mechanoreceptor enhancement:~Redox modulators for proprioceptive deficits.
\end{enumerate}

Novel sensory interfaces:

\begin{enumerate}
\def\labelenumi{\arabic{enumi}.}
\item
  Redox-coupled prosthetics: Devices that directly modulate O₂/O₂•⁻ equilibrium.
\item
  Magnetic field therapies: For sensory disorders via radical pair mechanism.
\item
  Thermal therapies: Precise temperature manipulation of redox thresholds.
\end{enumerate}

\section*{8. Summation}

The analyses presented herein demonstrate that neuronal electrical activity can be consistently interpreted as a manifestation of redox-mediated electronic dynamics rather than as a process fundamentally driven by transmembrane ionic flux. By integrating comparison with established models, quantitative predictions, and diverse experimental observations, the murburn framework emerges as a unified and chemically grounded description of excitability.

A key strength of the model lies in its predictive structure. Its appeal is mechanistic parsimony rather than freedom from fitting (the model still uses free or imposed parameters); the murburn formulation links measurable electrophysiological outputs: such as conduction velocity, waveform morphology, and threshold behavior; to physically interpretable variables including redox kinetics, transport efficiency, and environmental conditions. This enables direct experimental validation through perturbations in oxygen availability, ROS measurements, redox balance, solvent properties, ionic strength, isotope substitution, radical pair magnetic effects and external fields.

Importantly, the framework extends beyond neurons to a broader class of excitable systems, including cardiac tissue, photoreceptors, and artificial redox-active materials, suggesting that excitability is a general physicochemical phenomenon rooted in reaction--transport dynamics.

While the present work establishes the mid-scale dynamics of neuronal electricality, further developments are required to connect quantum-level electron transfer processes with macroscopic electrophysiological signals such as EEG and EMG. These extensions, along with targeted experimental tests, will determine the ultimate scope and applicability of the murburn paradigm.

\textbf{Disclaimers:} KMM wrote the first draft of the paper, presented the arguments and surveyed the literature to get the data. NS provided critical comments, identified lacunae, and corrected text. TH contributed with computational simulations/analyses of the murburn formulation and minor editorial revisions. MK provided crucial inputs. AA carried out the revision and consistency review of the manuscript and prepared the typeset version. This work was powered by Satyamjayatu: The Science \& Ethics Foundation.

\section*{References}
\begin{refslist}
Băng, G., \& Mellander, S. (1976). Effects of deuterium oxide on the electrical and mechanical properties of rat portal vein. Acta Physiologica Scandinavica, 96(1), 34--47. https://doi.org/10.1111/j.1748-1716.1976.tb10168.x

Brunori, M., Giuffrè, A., Nienhaus, K., Nienhaus, G. U., Vallone, B., \& Verzili, D. (2005). Neuroglobin, nitric oxide, and oxygen: Functional pathways and conformational changes. Proceedings of the National Academy of Sciences of the United States of America, 102(24), 8483--8488.

Chapman, R. A. (1967). The temperature dependence of the conduction velocity in the sensory nerves of the cat\textquotesingle s cornea. The Journal of Physiology, 189(1), 65--77. https://doi.org/10.1113/jphysiol.1967.sp008156

Conti, F. (1970). Nerve membrane polarization: The effect of temperature and of D₂O on the impedance of the membrane of the squid giant axon. Biophysical Journal, 10(3), 231--247. https://doi.org/10.1016/S0006-3495(70)86292-X

De Luccia, T. P. (2012). Mimosa pudica, Dionaea muscipula and anesthetics. Plant Signaling \& Behavior, 7(9), 1163--1167. https://doi.org/10.4161/psb.21000

Fisher, R. A. (1937). The wave of advance of advantageous genes. Annals of Eugenics, 7(4), 355--369.

FitzHugh, R. (1961). Impulses and physiological states in theoretical models of nerve membrane. Biophysical Journal, 1(6), 445--466.

Fourier, J. (1822). Théorie analytique de la chaleur. Firmin Didot.

Frankenhaeuser, B., \& Moore, L. E. (1963). The effect of temperature on the sodium and potassium permeability changes in myelinated nerve fibres of Xenopus laevis. The Journal of Physiology, 169(2), 431--437.

Ghodbane, S., Lahbib, A., Sakly, M., \& Abdelmelek, H. (2013). Bioeffects of static magnetic fields: Oxidative stress, genotoxic effects, and cancer studies. BioMed Research International, 2013, Article 602987. https://doi.org/10.1155/2013/602987

Ginzburg, V. L., \& Landau, L. D. (1950). On the theory of superconductivity. Zhurnal Eksperimental\textquotesingle noi i Teoreticheskoi Fiziki, 20, 1064--1082. (English translation: Men of Physics: L.D. Landau, Vol. I, Pergamon, 1965)

Hedrich, R., \& Kreuzer, I. (2023). Demystifying the Venus flytrap action potential. New Phytologist, 239(6), 2108--2112. https://doi.org/10.1111/nph.19113

Heimburg, T., \& Jackson, A. D. (2005). On soliton propagation in biomembranes and nerves. Proceedings of the National Academy of Sciences of the United States of America, 102(28), 9790--9795.

Hodgkin, A. L., \& Huxley, A. F. (1952). A quantitative description of membrane current and its application to conduction and excitation in nerve. The Journal of Physiology, 117(4), 500--544.

Hodgkin, A. L., \& Katz, B. (1949). The effect of temperature on the electrical activity of the giant axon of the squid. The Journal of Physiology, 109(1--2), 240--249. https://doi.org/10.1113/jphysiol.1949.sp004388

Hore, P. J. (2012). Are biochemical reactions affected by weak magnetic fields? Proceedings of the National Academy of Sciences of the United States of America, 109(4), 1357--1358. https://doi.org/10.1073/pnas.1120531109

Howarth, C., Gleeson, P., \& Attwell, D. (2012). Updated energy budgets for neural computation in the neocortex and cerebellum. Journal of Cerebral Blood Flow \& Metabolism, 32(7), 1222--1232. https://doi.org/10.1038/jcbfm.2012.35

Jacak, J. E., \& Jacak, W. A. (2020). New wave-type mechanism of saltatory conduction in myelinated axons and micro-saltatory conduction in C fibres. European Biophysics Journal, 49(5), 343--360. https://doi.org/10.1007/s00249-020-01442-z

Kolmogorov, A., Petrovskii, I., \& Piskunov, N. (1937). A study of the diffusion equation with increase in the amount of substance, and its application to a biological problem. Bulletin of Moscow State University, Mathematics and Mechanics, 1, 1--25.

Kourennyi, D. E., Liu, X., \& Barnes, S. (2004). Cyclic nucleotide modulation of neuronal voltage-gated calcium channels mediated by redox agents. The Journal of Neuroscience, 24(41), 9326--9331.

Manoj, K. M., Gideon, D. A., Bazhin, N. M., Tamagawa, H., Nirusimhan, V., Kavdia, M., \& Jaeken, L. (2023a). Na,K‐ATPase: A murzyme facilitating thermodynamic equilibriums at the membrane‐interface. Journal of Cellular Physiology, 238(1), 109--136. https://doi.org/10.1002/jcp.30925

Manoj, K. M., et al. (2026c). Structure and mechanism of cellular cation-transporters: Affinity-binding and murburn models. International Journal of Biological Macromolecules. https://doi.org/10.1016/j.ijbiomac.2026.150614

McFadden, J. (2002). The conscious electromagnetic information (Cemi) field theory: The hard problem made easy? Journal of Consciousness Studies, 9(8), 45--60.

McLean, M. J., Engström, S., Holcomb, R. R., \& Sanchez, D. (2003). A static magnetic field modulates severity of audiogenic seizures and anticonvulsant effects of phenytoin in DBA/2 mice. Epilepsy Research, 55(1--2), 105--116. https://doi.org/10.1016/S0920-1211(03)00114-0

Moore, J. W. (1969). Ionic conductance changes in voltage clamped crayfish axons bathed in deuterium oxide. The Journal of General Physiology, 54(3), 334--350.

Mukandala, G., Tynan, R., Lanigan, S., \& O\textquotesingle Connor, J. J. (2016). The effects of hypoxia and inflammation on synaptic signaling in the CNS. Brain Sciences, 6(1), 6. https://doi.org/10.3390/brainsci6010006

Nagumo, J., Arimoto, S., \& Yoshizawa, S. (1962). An active pulse transmission line simulating nerve axon. Proceedings of the IRE, 50(10), 2061--2070.

Paintal, A. S. (1965). Effects of temperature on conduction in single vagal and saphenous myelinated nerve fibres of the cat. The Journal of Physiology, 180(1), 20--49. https://doi.org/10.1113/jphysiol.1965.sp007686

Pockett, S. (2000). The nature of consciousness. iUniverse.

Przybylski, A. T., \& Fox, S. W. (1984). Excitable artificial cells of proteinoid. Applied Biochemistry and Biotechnology, 10, 301--307. https://doi.org/10.1007/BF02783764

Przybylski, A. T., Stratten, W. P., Syren, R. M., \& Fox, S. W. (1982). Membrane, action, and oscillatory potentials in simulated protocells. Naturwissenschaften, 69(12), 561--563. https://doi.org/10.1007/BF00396351

Rall, W. (1962). Theory of physiological properties of dendrites. Annals of the New York Academy of Sciences, 96, 1071--1092.

Ritchie, J. M., \& Straub, R. W. (1980). The oxygen consumption of mammalian non-myelinated nerve fibres at rest and during activity. The Journal of Physiology, 304, 109--121. https://doi.org/10.1113/jphysiol.1980.sp013316

Rodgers, C. T. (2009). Magnetic field effects in chemical systems. Pure and Applied Chemistry, 81(1), 19--43. https://doi.org/10.1351/PAC-CON-08-10-18

Rosen, A. D. (1996). Inhibition of calcium channel activation in GH3 cells by static magnetic fields. *Biochimica et Biophysica Acta (BBA) - Biomembranes, 1282*(1), 149--155. https://doi.org/10.1016/0005-2736(96)00052-0

Rosen, A. D. (2003). Mechanism of action of moderate-intensity static magnetic fields on biological systems. Cell Biochemistry and Biophysics, 39(2), 163--173. https://doi.org/10.1385/CBB:39:2:163

Sibaoka, T. (1991). Rapid plant movements triggered by action potentials. The Botanical Magazine (Tokyo), 104, 73--95. https://doi.org/10.1007/BF02493405

Spyropoulos, C. S. (1957). The effects of deuterium oxide on the properties of nerve fibres. The Journal of General Physiology, 40(6), 849--857. https://doi.org/10.1085/jgp.40.6.849

Takahashi, K.-I., \& Copenhagen, D. R. (1996). The effects of reactive oxygen species on the membrane potential of retinal horizontal cells. Journal of Neurophysiology, 75(1), 191--196.

Thomson, W. (Lord Kelvin). (1855). On the theory of the electric telegraph. Proceedings of the Royal Society.

Turing, A. M. (1952). The chemical basis of morphogenesis. Philosophical Transactions of the Royal Society of London B, 237(641), 37--72.

Volkov, A. G., Foster, J. C., \& Markin, V. S. (2010). Signal transduction in Mimosa pudica: Biologically closed electrical circuits. Plant, Cell \& Environment, 33(5), 816--827. https://doi.org/10.1111/j.1365-3040.2009.02108.x

Wakasugi, K., Nakano, T., Kitatsuji, C., \& Morishima, I. (2003). Oxidized human neuroglobin acts as a heterotrimeric Gα protein guanine nucleotide dissociation inhibitor. Journal of Biological Chemistry, 278(38), 36505--36512.

Zhang, X., Yarema, K., \& Xu, A. (2016). Biological effects of static magnetic fields (Chapter 7). Springer.

\end{refslist}

\clearpage
\section*{Supplementary Information}
This Supplementary Information collects the expected outcomes across stimuli (Tables~A--F) and the Python-based simulation outputs for the murburn model (Figures~S1--S4).

\textbf{Table A: Effect of temperature}

\begin{longtable}[]{@{}
  >{\raggedright\arraybackslash}p{(\columnwidth - 10\tabcolsep) * \real{0.1460}}
  >{\raggedright\arraybackslash}p{(\columnwidth - 10\tabcolsep) * \real{0.1545}}
  >{\raggedright\arraybackslash}p{(\columnwidth - 10\tabcolsep) * \real{0.1652}}
  >{\raggedright\arraybackslash}p{(\columnwidth - 10\tabcolsep) * \real{0.1653}}
  >{\raggedright\arraybackslash}p{(\columnwidth - 10\tabcolsep) * \real{0.1975}}
  >{\raggedright\arraybackslash}p{(\columnwidth - 10\tabcolsep) * \real{0.1716}}@{}}
\toprule\noalign{}
\begin{minipage}[b]{\linewidth}\raggedright
\textbf{Aspect}
\end{minipage} & \begin{minipage}[b]{\linewidth}\raggedright
\textbf{CMPT Expectation}
\end{minipage} & \begin{minipage}[b]{\linewidth}\raggedright
\textbf{Murburn Expectation}
\end{minipage} & \begin{minipage}[b]{\linewidth}\raggedright
\textbf{Experimental Observations}
\end{minipage} & \begin{minipage}[b]{\linewidth}\raggedright
\textbf{Documented works}
\end{minipage} & \begin{minipage}[b]{\linewidth}\raggedright
\textbf{Contextual Interpretation}
\end{minipage} \\
\midrule\noalign{}
\endhead
\bottomrule\noalign{}
\endlastfoot
NCV vs Temperature & NCV modestly increases with temperature due to faster channel kinetics & NCV moderately increases due to enhanced redox dynamics, with a peak then decline at extremes & Nerve conduction studies show NCV increases with temperature (warmer limbs show faster NCV) & Clinical standard texts; thermally induced NCV changes well documented & Temperature influences both ionic gating and redox chemical rates; bell-shape at extremes due to protein stability \\
Amplitude / Waveform & AP amplitude usually unchanged within physiological range & AP amplitude sensitive at extremes due to redox sensibility & Mild cooling slows conduction and can widen AP duration & Patient electrophysiology labs show slowed NCV and prolonged latencies with cooling & Within physiological T range, both frameworks compatible, but redox sensitivity better predicts heat-shock phenomena \\
Failure at extreme T & Conduction blocks at very low or high T & Early conduction failure predicted due to redox destabilization & Extreme cooling freezes ion channels and blocks AP & Practicals using in vitro cooling experiments (textbook labs) & Both frameworks accept conduction blocks, but underlying causal mechanisms differ \\
\end{longtable}

References: https://www.sciencedirect.com/topics/neuroscience/sensory-nerve-conduction

\textbf{Table B: Effect of solvent (protons/pH/buffering/D\textsubscript{2}O)}

\begin{longtable}[]{@{}
  >{\raggedright\arraybackslash}p{(\columnwidth - 10\tabcolsep) * \real{0.1481}}
  >{\raggedright\arraybackslash}p{(\columnwidth - 10\tabcolsep) * \real{0.1585}}
  >{\raggedright\arraybackslash}p{(\columnwidth - 10\tabcolsep) * \real{0.1607}}
  >{\raggedright\arraybackslash}p{(\columnwidth - 10\tabcolsep) * \real{0.1666}}
  >{\raggedright\arraybackslash}p{(\columnwidth - 10\tabcolsep) * \real{0.1934}}
  >{\raggedright\arraybackslash}p{(\columnwidth - 10\tabcolsep) * \real{0.1727}}@{}}
\toprule\noalign{}
\begin{minipage}[b]{\linewidth}\raggedright
\textbf{Aspect}
\end{minipage} & \begin{minipage}[b]{\linewidth}\raggedright
\textbf{CMPT Expectation}
\end{minipage} & \begin{minipage}[b]{\linewidth}\raggedright
\textbf{Murburn Expectation}
\end{minipage} & \begin{minipage}[b]{\linewidth}\raggedright
\textbf{Experimental Observations}
\end{minipage} & \begin{minipage}[b]{\linewidth}\raggedright
\textbf{Documented works}
\end{minipage} & \begin{minipage}[b]{\linewidth}\raggedright
\textbf{Contextual Interpretation}
\end{minipage} \\
\midrule\noalign{}
\endhead
\bottomrule\noalign{}
\endlastfoot
Overshoot / AP amplitude & Generally unaffected by simple solvents & Sensitive to proton mobility and redox environment, altering attack/decay & D₂O experiments show changes in AP kinetics and amplitude & Neurophysiology experimentation (D₂O substitution alters excitability) & Protonic mobility affects electron--proton coupling; not explained by pure ion gating \\
Spike symmetry & Unchanged with buffer/D₂O & Distorted due to altered proton coupling to redox mediators & Observed kinetic changes in AP with solvent isotopic substitution & Multiple classical physiology labs have noted D₂O effects on excitability & Buffering and solvent effects imply deeper chemical coupling than channels alone \\
\end{longtable}

\textbf{Table C: Effect of ionic strength}

\begin{longtable}[]{@{}
  >{\raggedright\arraybackslash}p{(\columnwidth - 10\tabcolsep) * \real{0.1067}}
  >{\raggedright\arraybackslash}p{(\columnwidth - 10\tabcolsep) * \real{0.1554}}
  >{\raggedright\arraybackslash}p{(\columnwidth - 10\tabcolsep) * \real{0.1607}}
  >{\raggedright\arraybackslash}p{(\columnwidth - 10\tabcolsep) * \real{0.1663}}
  >{\raggedright\arraybackslash}p{(\columnwidth - 10\tabcolsep) * \real{0.1554}}
  >{\raggedright\arraybackslash}p{(\columnwidth - 10\tabcolsep) * \real{0.2555}}@{}}
\toprule\noalign{}
\begin{minipage}[b]{\linewidth}\raggedright
\textbf{Aspect}
\end{minipage} & \begin{minipage}[b]{\linewidth}\raggedright
\textbf{CMPT Expectation}
\end{minipage} & \begin{minipage}[b]{\linewidth}\raggedright
\textbf{Murburn Expectation}
\end{minipage} & \begin{minipage}[b]{\linewidth}\raggedright
\textbf{Experimental Observations}
\end{minipage} & \begin{minipage}[b]{\linewidth}\raggedright
\textbf{Documented works}
\end{minipage} & \begin{minipage}[b]{\linewidth}\raggedright
\textbf{Contextual Interpretation}
\end{minipage} \\
\midrule\noalign{}
\endhead
\bottomrule\noalign{}
\endlastfoot
NCV vs Ionic Strength & Slight monotonic effect or saturation due to electrolyte support & Bell-shaped curve: rising up to an optimum and then decline due to over-shielding & Lobster giant axon studies, increased {[}K⁺{]}o progressively impede conduction & Peacock et al. (rat spinal CAP work) & Ionic strength modulates conductance indirectly but does affect propagation speed \\
Spike width & Minimal effect & Optimized near physiologic ionic strength; broader/failed at extremes & Increased extracellular K⁺ attenuates spike amplitude & Rat spinal nerve CAP data & Classical view: ion gradients matter, but the bell shape fits murburn's shielding hypothesis \\
Failure at very high I & Unlikely within physiological range & Predicted due to excessive screening & Conduction block with high extracellular K⁺ & Peacock et al. & Supports screening/redistribution effects beyond simple ions \\
\end{longtable}

References: Peacock JM, Orchardson R. Effects of potassium ions on action potential conduction in A- and C-fibers of rat spinal nerves. J Dent Res. 1995 Feb;74(2):634-41. doi: 10.1177/00220345950740020301.

https://nba.uth.tmc.edu/neuroscience/m/s1/chapter02.html

\textbf{Table D: Effect of oxygen}

\begin{longtable}[]{@{}
  >{\raggedright\arraybackslash}p{(\columnwidth - 10\tabcolsep) * \real{0.1510}}
  >{\raggedright\arraybackslash}p{(\columnwidth - 10\tabcolsep) * \real{0.1598}}
  >{\raggedright\arraybackslash}p{(\columnwidth - 10\tabcolsep) * \real{0.1635}}
  >{\raggedright\arraybackslash}p{(\columnwidth - 10\tabcolsep) * \real{0.1974}}
  >{\raggedright\arraybackslash}p{(\columnwidth - 10\tabcolsep) * \real{0.1556}}
  >{\raggedright\arraybackslash}p{(\columnwidth - 10\tabcolsep) * \real{0.1727}}@{}}
\toprule\noalign{}
\begin{minipage}[b]{\linewidth}\raggedright
\textbf{Aspect}
\end{minipage} & \begin{minipage}[b]{\linewidth}\raggedright
\textbf{CMPT Expectation}
\end{minipage} & \begin{minipage}[b]{\linewidth}\raggedright
\textbf{Murburn Expectation}
\end{minipage} & \begin{minipage}[b]{\linewidth}\raggedright
\textbf{Experimental Observations}
\end{minipage} & \begin{minipage}[b]{\linewidth}\raggedright
\textbf{References}
\end{minipage} & \begin{minipage}[b]{\linewidth}\raggedright
\textbf{Contextual Interpretation}
\end{minipage} \\
\midrule\noalign{}
\endhead
\bottomrule\noalign{}
\endlastfoot
Resting Vm & Stable until significant ATP loss & Stable until redox systems are disrupted & Neurons maintain Vm initially under mild hypoxia & Hypoxia neuronal studies & Both frameworks allow initial stability \\
AP Overshoot & Largely stable until metabolic failure & Early drop due to redox disruption & Hypoxia reduces AP amplitude \& can depress excitability & Chronic hypoxia / perfusion review & Early effects suggest redox involvement beyond ATP depletion \\
NCV & Stable until ATP is depleted & Early reduction due to oxygen/redox depletion & Hypoxia reduces synaptic/neuronal activity & Mukandala et al. review & Hypoxia effects often begin before major ATP fall \\
\end{longtable}

References: Nieber K. Hypoxia and neuronal function under in vitro conditions. Pharmacol Ther. 1999 Apr;82(1):71-86. doi: 10.1016/s0163-7258(98)00061-8.

\textbf{Table E: Effect of myelination/geometry}

\begin{longtable}[]{@{}
  >{\raggedright\arraybackslash}p{(\columnwidth - 10\tabcolsep) * \real{0.1588}}
  >{\raggedright\arraybackslash}p{(\columnwidth - 10\tabcolsep) * \real{0.1607}}
  >{\raggedright\arraybackslash}p{(\columnwidth - 10\tabcolsep) * \real{0.1607}}
  >{\raggedright\arraybackslash}p{(\columnwidth - 10\tabcolsep) * \real{0.1666}}
  >{\raggedright\arraybackslash}p{(\columnwidth - 10\tabcolsep) * \real{0.1571}}
  >{\raggedright\arraybackslash}p{(\columnwidth - 10\tabcolsep) * \real{0.1961}}@{}}
\toprule\noalign{}
\begin{minipage}[b]{\linewidth}\raggedright
\textbf{Aspect}
\end{minipage} & \begin{minipage}[b]{\linewidth}\raggedright
\textbf{CMPT Expectation}
\end{minipage} & \begin{minipage}[b]{\linewidth}\raggedright
\textbf{Murburn Expectation}
\end{minipage} & \begin{minipage}[b]{\linewidth}\raggedright
\textbf{Experimental Observations}
\end{minipage} & \begin{minipage}[b]{\linewidth}\raggedright
\textbf{References}
\end{minipage} & \begin{minipage}[b]{\linewidth}\raggedright
\textbf{Contextual Interpretation}
\end{minipage} \\
\midrule\noalign{}
\endhead
\bottomrule\noalign{}
\endlastfoot
NCV vs Myelination & Myelin greatly increases NCV via insulation & Myelin enhances coherent relay and reduces radial dissipation & Myelinated axons conduct much faster than unmyelinated & Saltatory conduction wiki; myelinated fiber sources & Both models recognize geometric/myelin dependences \\
Internodal Distance & Optimal spacing gives max NCV & Also optimal for stable redox relay & Nodes of Ranvier spacing correlated with fast conduction & Saltatory conduction mechanics & Same geometry effects but different causal logic \\
\end{longtable}

References: https://en.wikipedia.org/wiki/Saltatory\_conduction

\textbf{Table F: Effect of redox additives}

\begin{longtable}[]{@{}
  >{\raggedright\arraybackslash}p{(\columnwidth - 10\tabcolsep) * \real{0.1315}}
  >{\raggedright\arraybackslash}p{(\columnwidth - 10\tabcolsep) * \real{0.1527}}
  >{\raggedright\arraybackslash}p{(\columnwidth - 10\tabcolsep) * \real{0.1527}}
  >{\raggedright\arraybackslash}p{(\columnwidth - 10\tabcolsep) * \real{0.2463}}
  >{\raggedright\arraybackslash}p{(\columnwidth - 10\tabcolsep) * \real{0.1440}}
  >{\raggedright\arraybackslash}p{(\columnwidth - 10\tabcolsep) * \real{0.1727}}@{}}
\toprule\noalign{}
\begin{minipage}[b]{\linewidth}\raggedright
\textbf{Aspect}
\end{minipage} & \begin{minipage}[b]{\linewidth}\raggedright
\textbf{CMPT Expectation}
\end{minipage} & \begin{minipage}[b]{\linewidth}\raggedright
\textbf{Murburn Expectation}
\end{minipage} & \begin{minipage}[b]{\linewidth}\raggedright
\textbf{Experimental Observations}
\end{minipage} & \begin{minipage}[b]{\linewidth}\raggedright
\textbf{References}
\end{minipage} & \begin{minipage}[b]{\linewidth}\raggedright
\textbf{Contextual Interpretation}
\end{minipage} \\
\midrule\noalign{}
\endhead
\bottomrule\noalign{}
\endlastfoot
AP amplitude & No direct expectation & Sensitive to redox state & Antioxidants modulate excitability in some studies & Cameron antioxidant nerve work & Effects imply redox involvement not accounted for by ionic view \\
NCV & No direct expectation & Bell-shaped dependence on redox species levels & Hypoxia/inflammation affect conduction & Mukandala review & Redox environment affects conduction independent of ion pumps \\
\end{longtable}

Figures S1--S4 present Python-based simulation outputs of the murburn framework, reconstructed from the Γ-relaxation logic underlying the theoretical framework of Part 1. These support the predictions summarised in Tables A--F above

\begin{longtable}[]{@{}
  >{\raggedright\arraybackslash}p{(\columnwidth - 0\tabcolsep) * \real{1.0000}}@{}}
\toprule\noalign{}
\begin{minipage}[b]{\linewidth}\raggedright
\includegraphics[width=6.26806in,height=4.47708in]{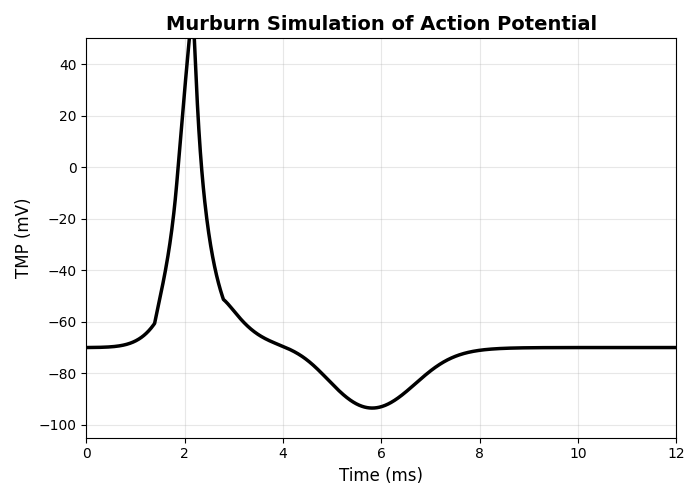}

\textbf{Figure S1. Murburn simulation of the action potential waveform.} The transmembrane potential (TMP) is obtained from the murburn EHP relaxation equation: ∂φ/∂t = (Γ(t) − φ)/τ, where Γ(t) encodes the biphasic EDP--ESP redox program. The resting TMP which is around −70 mV arises from the stable fixed point of the local redox dynamics. Following threshold perturbation, the EDP phase drives φ downward (lower electron-holding capacity), producing rapid depolarization to +47 mV. The ESP phase subsequently elevates Γ above φ\_rest, driving after-hyperpolarization to ≈ −95 mV before recovery to rest. The simulated waveform reproduces all characteristic phases of the canonical action potential such as the resting state, depolarization, peak, repolarization, after-hyperpolarization and recovery. This can be reproduce using a single murburn Γ-relaxation framework, without requiring explicit ion-channel gating variables. The parameters used were: τ\_local = 0.12 ms for the EDP-dominated phase, τ\_bulk = 0.4 ms for the recovery phase, with EDP and ESP amplitudes of 14.2 and 22.0 φ-units, respectively.
\end{minipage} \\
\midrule\noalign{}
\endhead
\bottomrule\noalign{}
\endlastfoot
\end{longtable}

\begin{longtable}[]{@{}
  >{\raggedright\arraybackslash}p{(\columnwidth - 0\tabcolsep) * \real{1.0000}}@{}}
\toprule\noalign{}
\begin{minipage}[b]{\linewidth}\raggedright
\includegraphics[width=6.26806in,height=3.76111in]{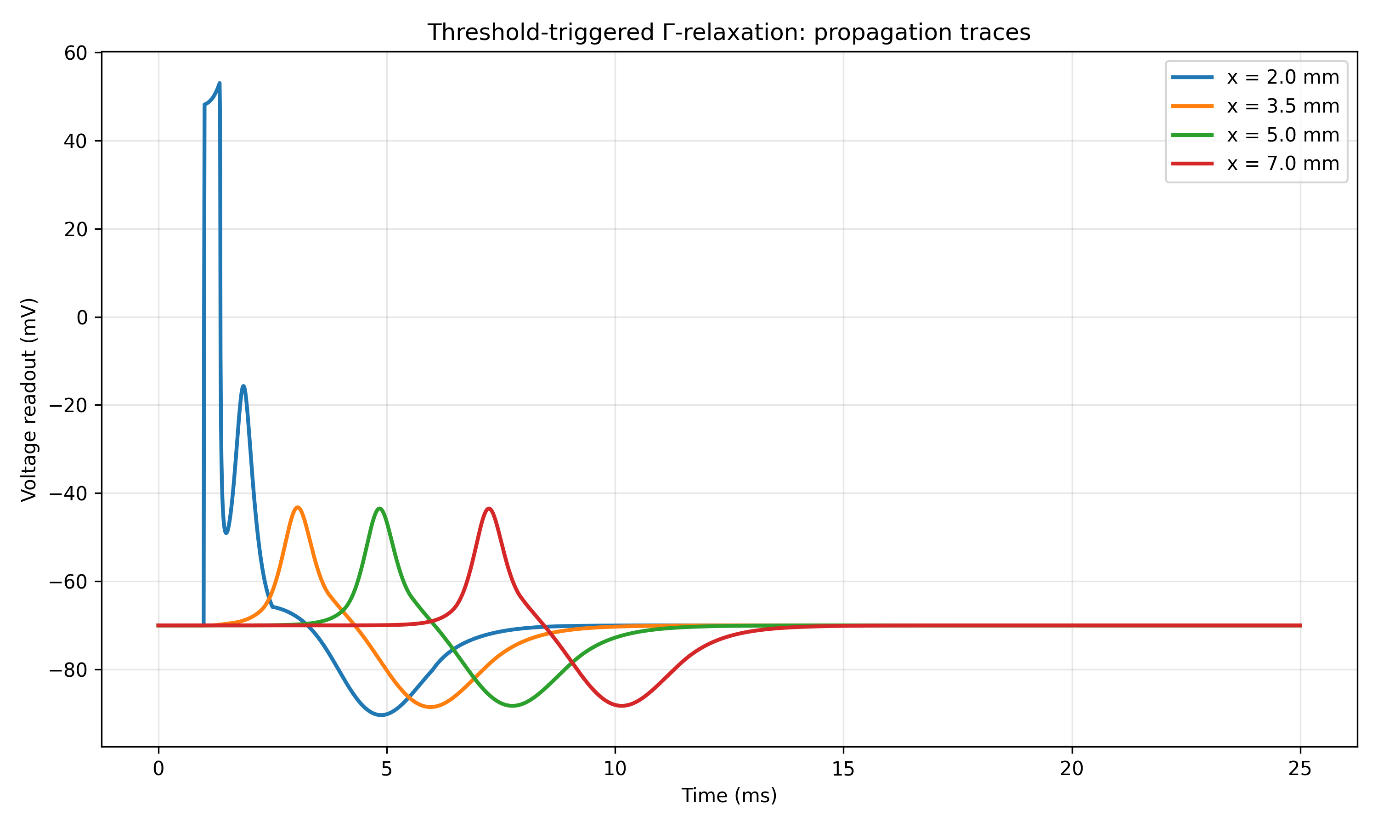}

\textbf{Figure S2. Propagating action potential waveforms at four axonal positions.} The spatial Γ-relaxation PDE (∂φ/∂t = D∂²φ/∂x² + (Γ(x,t) − φ)/τ) was solved on a 10 mm axon (NX = 300 grid points) with a threshold-triggered local Γ-program. A depolarising stimulus was applied at x = 2.0 mm (t = 1.0--1.35 ms). The voltage traces in this figure are presented at four spatial positions, the stimulus location x = 2.0 mm and three downstream points x = 3.5, 5.0, and 7.0 mm. As the signal propagates, the waveform conserves both its amplitude and characteristic shape, consistent with a self-sustaining regenerative redox wave rather than passive diffusion. The progressive delay in peak arrival across spatial positions produces a measurable conduction velocity (NCV ≈ 0.8 m/s under baseline conditions). The after-hyperpolarization trough (ESP phase) also propagates with high fidelity, indicating effective enforcement of the refractory phase at each location. This directional, shape-preserving propagation represents the spatial manifestation of the all-or-none behavior discussed in Section 4 of the main text.
\end{minipage} \\
\midrule\noalign{}
\endhead
\bottomrule\noalign{}
\endlastfoot
\end{longtable}

\begin{longtable}[]{@{}
  >{\raggedright\arraybackslash}p{(\columnwidth - 0\tabcolsep) * \real{1.0000}}@{}}
\toprule\noalign{}
\begin{minipage}[b]{\linewidth}\raggedright
\includegraphics[width=6.26806in,height=3.41875in]{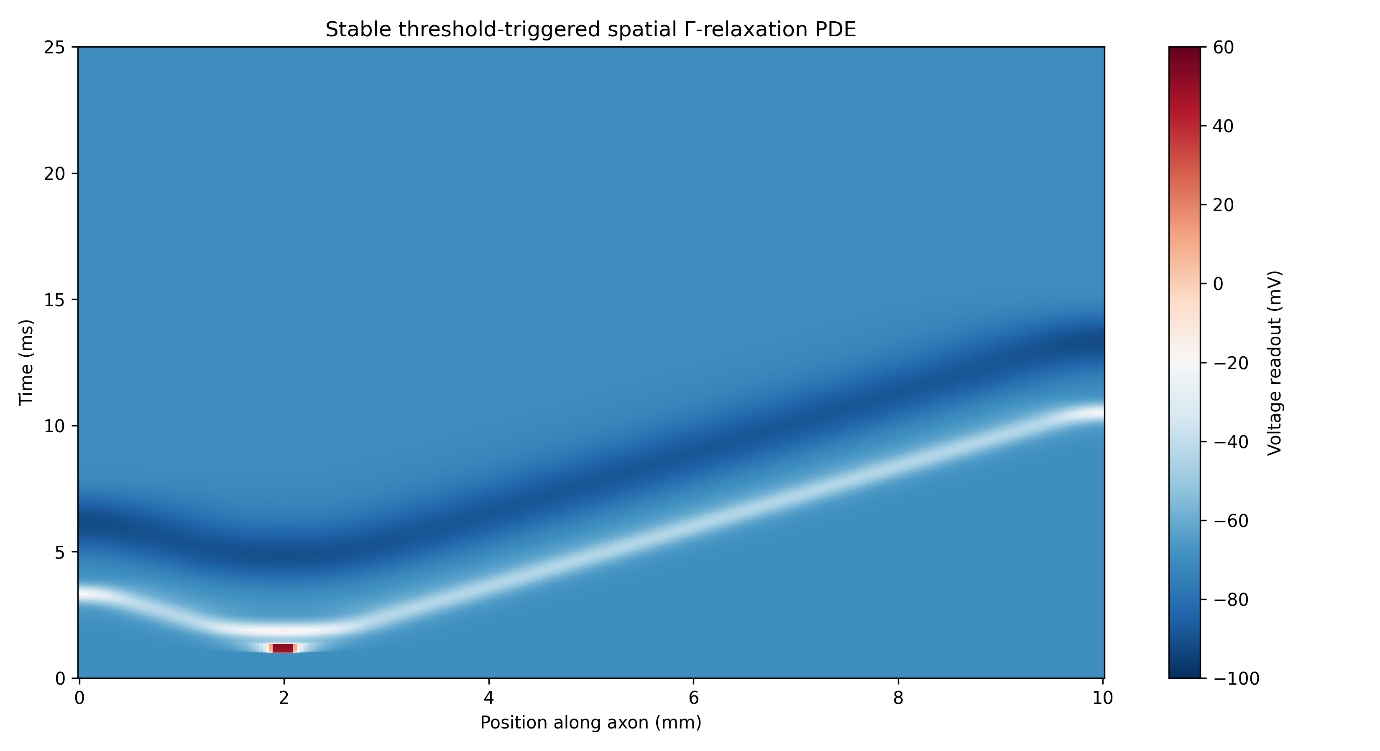}

\textbf{Figure S3. Spatiotemporal voltage map of murburn axonal propagation.} Colour encodes the voltage readout V = −(RT/F)ln φ across the full axon length (0--10 mm) and simulation duration (0--25 ms). The localised red region at x = 2 mm, t = 1 ms marks the stimulus-triggered EDP event. The propagating wave appears as a diagonal light band (white = near-rest, blue = hyperpolarised) moving at constant velocity from left to right, confirming a true traveling wave solution to the murburn PDE. The slope of this band directly corresponds to the NCV: Δx/Δt ≈ 0.8 m/s. The diffuse dark-blue region at the stimulus site (t = 3--7 ms) represents the after-hyperpolarisation and local refractory period, during which no second wave can be initiated. The absence of signal at x \textless{} 2 mm confirms unidirectional propagation consistent with the refractory logic. The constant slope of the propagating band across the full axon length validates the analytical NCV expression v = 2√(Dσ/τ) derived in Part 1.
\end{minipage} \\
\midrule\noalign{}
\endhead
\bottomrule\noalign{}
\endlastfoot
\end{longtable}

\begin{longtable}[]{@{}
  >{\raggedright\arraybackslash}p{(\columnwidth - 0\tabcolsep) * \real{1.0000}}@{}}
\toprule\noalign{}
\begin{minipage}[b]{\linewidth}\raggedright
\includegraphics[width=5.98108in,height=5.58216in]{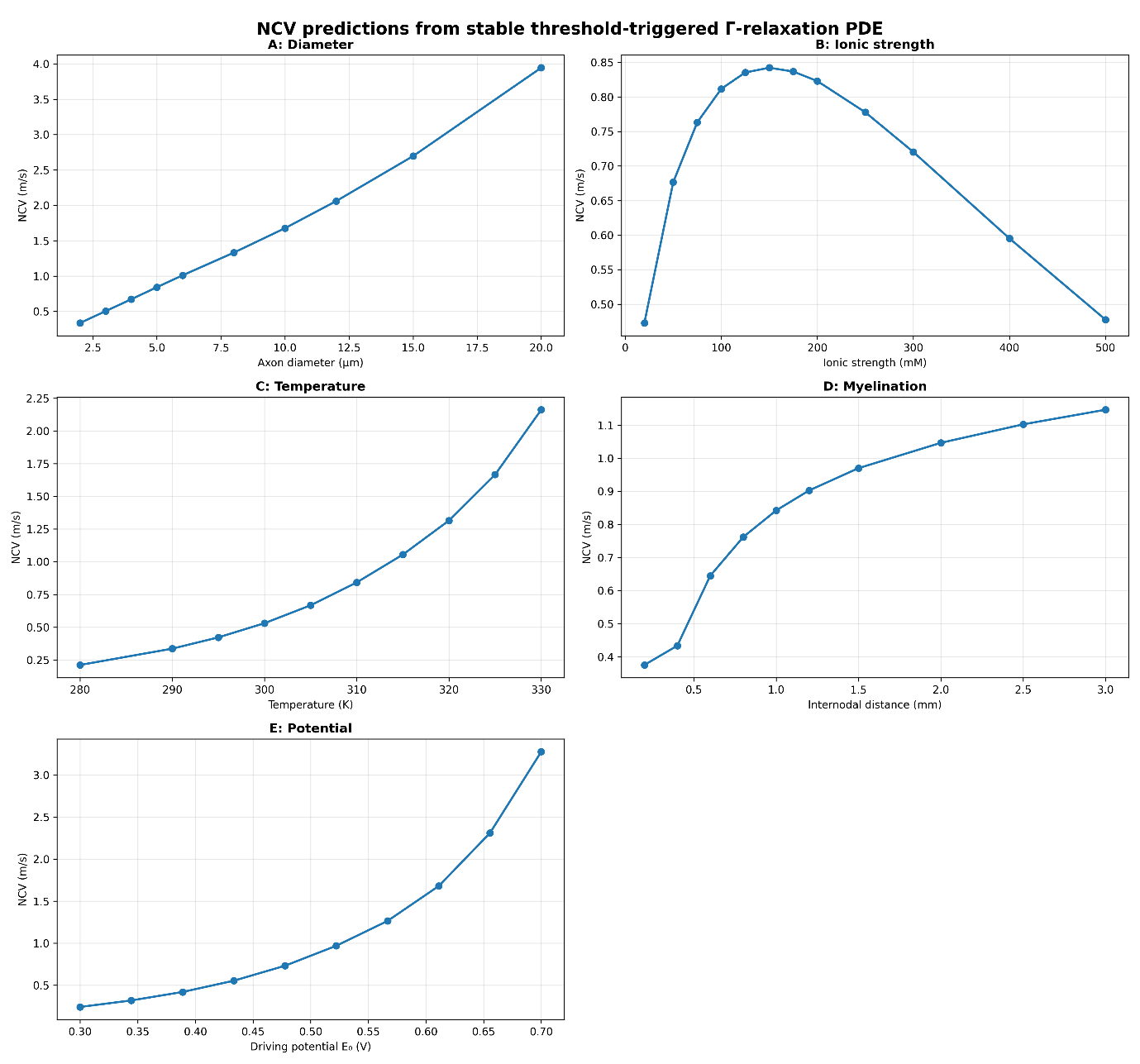}

\textbf{Figure S4. Predicted NCV dependence on five experimental parameters from the murburn spatial PDE.} Conduction velocity was measured from the first-trigger time at two axonal positions (x = 30\% and 70\% of total length) across parameter sweeps. (A) Axon diameter: NCV scales as d² via the transport coefficient D ∝ d², matching the empirical diameter--velocity relationship. (B) Ionic strength: NCV exhibits a bell-shaped dependence peaking near 125--150 mM (physiological), arising from the Debye--Hückel screening of redox-active species, a distinctive murburn prediction not accommodated by HH-cable monotonic conductance dependence. (C) Temperature: NCV increases monotonically with temperature, consistent with Q₁₀ = 2--2.5 for redox/electron-transfer processes (cf. Table 5, main text), substantially exceeding the Q₁₀ = 1.1 expected for diffusion-limited processes. (D) Myelination (internodal distance): NCV increases with internodal spacing, consistent with enhanced longitudinal redox relay coherence in myelinated segments. (E) Driving redox potential E₀: NCV shows exponential sensitivity to the electrochemical driving potential and reflect the Arrhenius-like dependence of electron-transfer rates. All five plots were generated from the same PDE framework, thereby demonstrating that he murburn model produces qualitatively and quantitatively distinct predictions across diverse physiological perturbations from a single unified equation.
\end{minipage} \\
\midrule\noalign{}
\endhead
\bottomrule\noalign{}
\endlastfoot
\end{longtable}

\end{document}